\documentclass[aps,physrev,amsmath,amssymb,notitlepage,twocolumn,longbibliography,reprint]{revtex4-2}
\usepackage{amsmath,epsfig,amssymb}
\usepackage{bbm}
\usepackage{times}
\usepackage{color}
\usepackage{amsthm}
\usepackage{color}
\usepackage[caption=false]{subfig}
\usepackage{multirow}
\usepackage[normalem]{ulem}
\usepackage{qcircuit}
\usepackage{braket}
\usepackage{amsfonts,amssymb,dsfont}
\usepackage{graphicx}
\usepackage{soul}
\usepackage{hyperref}
\hypersetup{
    colorlinks=true,       
    linkcolor=red,          
  citecolor=magenta,        
    filecolor=magenta,      
    urlcolor=cyan,           
    runcolor=cyan
}

\usepackage[capitalise,compress]{cleveref}
\crefname{section}{Sec.}{Secs.}
\Crefname{section}{Section}{Sections}

\newcommand{\todo}[1]{}
\renewcommand{\todo}[1]{{\color{red} TODO: {#1}}}
\newcommand{\question}[1]{}
\renewcommand{\question}[1]{{\color{red} QUESTION: {#1}}}

\DeclareMathOperator{\sinc}{sinc}

\begin{document} 
%
\title{SchWARMA: A model-based approach for time-correlated noise in quantum circuits}
\author{Kevin Schultz, Gregory Quiroz, Paraj Titum, and B.D. Clader}
\date{\today}
\affiliation{
 Johns Hopkins University Applied Physics Laboratory\\
 11100 Johns Hopkins Road, Laurel, MD, 20723, USA
}

\begin{abstract}

    Temporal noise correlations are ubiquitous in quantum systems, yet often neglected in the analysis of quantum circuits due to the complexity required to accurately characterize and model them. Autoregressive moving average (ARMA) models are a well-known technique from time series analysis that model time correlations in data.  By identifying the space of completely positive trace preserving (CPTP) quantum operations with a particular matrix manifold, we 
    generalize ARMA models to the space of CPTP maps to parameterize and simulate temporally correlated noise in quantum circuits. This approach, denoted Schr\"{o}dinger Wave ARMA (SchWARMA), provides a natural path for generalization of classic techniques from signal processing, control theory, and system identification for which ARMA models and linear systems are essential. This enables the broad theory of classical signal processing to be applied to quantum system simulation, characterization, and noise mitigation.

\end{abstract}

\maketitle

\section{Introduction}\label{sec:intro}
The circuit model of quantum computing \cite{deutsch1985quantum, Yao1993, nielsen2010quantum} has been used extensively to prove algorithmic speedups \cite{deutsch1992rapid, Grover1996, Shor1997} as well as the theory of fault-tolerant quantum computing \cite{Shor1996, Aharonov1997, PhysRevA.57.127}. This model makes an implicit assumption that a quantum computation can be broken up into a series of (possibly imperfect) incremental gates. 
This assumption generally implies that noise processes are uncorrelated in space and time. 
However, the presence of time-correlated noise, which is the focus of this Article, is well established in actual physical systems, e.g., $1/f^\alpha$ noise in superconducting qubits \cite{PhysRevLett.99.187006, Bylander:2011wc, Burnett:2014tv, PhysRevApplied.6.041001, Burnett:2019ti}. 

Characterization of time-correlated noise processes typically involves non-parametric reconstructions of the entire noise spectra \cite{PhysRevLett.107.230501,  Bylander:2011wc, PhysRevA.98.013414, PhysRevLett.114.017601, PhysRevLett.120.260504, PhysRevApplied.10.044017, Frey2017ion} 
requiring a large number of experiments. Furthermore, outside of a few closed-form cases, there is no more efficient way of forward simulating the dynamics of time-correlated noise 
short of brute-force numerical integration of the stochastic Liouville equation \cite{kubo1963stochastic} for the entire circuit.  The time step for such a simulation is typically orders of magnitude smaller than the nominal ``gate duration,'' resulting in a large number of matrix operations 
per gate. This approach is computationally very expensive, which has prevented much in the way of analysis of time-correlations in quantum circuits beyond a few qubits. 

Understanding the impact of time-correlated errors on quantum computation is important as they are known to have an impact on quantum error correction. These errors have been considered in the context of proving fault-tolerant threshold theorems \cite{PhysRevLett.96.050504, PhysRevA.71.012336, Aliferis2006}, although exactly how well quantum error correction will suppress noise in the presence of these errors is not well understood, but we do know of some examples where their presence can lead to fairly detrimental affects \cite{clader2021impact}.

In this Article, we introduce a model of temporally correlated noise that allows the wide-ranging field of classical time-series analysis to be applied to noisy circuit simulation and quantum noise spectroscopy. Our model draws on the technique of autoregressive moving average (ARMA) models \cite{whittle1963prediction,box2015time}, linking open quantum systems to discrete time linear systems theory and model-based spectrum estimation \cite{kay1981spectrum}. With simulation and estimation of noisy quantum dynamics in mind, we generalize ARMA to the space of completely positive trace preserving (CPTP) maps, which we denote Schr\"{o}dinger Wave ARMA (SchWARMA).
We convey the efficacy and versatility of this method through a set of examples of temporally correlated noise-affected dynamics from quantum control and error correction.

SchWARMA has a natural connection to semi-classical stochastic Hamiltonian noise in quantum systems, but is more general as it can also model correlated errors in \textit{dissipative} system-baths as in a stochastic form of Lindblad master equation \cite{breuer2002theory}.  In this sense, the errors considered  here are fundamentally Markovian with respect to any quantum bath, as the dynamics produced are correlated yet remain completely positive (CP) divisible \cite{rivas2014quantum}.  
Techniques such as \cite{cerrillo2014non} that seek to model non-Markovian quantum dynamics are fundamentally different from this noisy-circuit approach, as non-CP error channels interspersed with perfect gates can produce non-physical states.  That said, our technique can be generalized to simulate system-bath models of non-Markovian quantum dynamics at an obvious reduction of the number of qubits simulated, but we leave this to future work. 

\section{The SchWARMA Model}\label{sec:model}
An ARMA model relates the output $y_k$, at discrete steps $k$, to a series of inputs $x_k$  and prior outputs via~\cite{whittle1963prediction,box2015time},
\begin{equation}\label{eq:ARMA}
   y_k = \underbrace{\sum_{i=1}^p a_{i}
    y_{k-i}}_{AR}+\underbrace{\sum_{j=0}^{q} b_{j}x_{k-j}}_{MA}\,.
\end{equation}
The set $\{a_i\}$ defines the autoregressive portion of the model, and $\{b_j\}$ the moving average portion with $p$ and $q+1$ elements of each set respectively. 
The power spectrum $S_y(\omega)$ of the output $y$ of an ARMA model driven by i.i.d. Gaussian noise is 
\begin{equation}
    S_y(\omega)=\frac{\left|\sum_{k=0}^q  b_k \exp(-ik\omega)\right|^2}{\left|1+\sum_{k=1}^pa_k\exp(-ik\omega)\right|^2}\,,
\end{equation}
and ARMA models can approximate \textit{any} discrete-time power spectrum to arbitrary accuracy \cite{holan2010arma}.
%
ARMA models (and their many generalizations \cite{holan2010arma}) are commonly used in classical time series analysis to parametrically define, estimate, and simulate time-correlated processes \cite{box2015time}, and  have wide-ranging applications all the way from actuarial science \cite{ives2010analysis} to zoology \cite{frees1997forecasting}. 


 We generalize ARMA models to CPTP maps by using \textit{Stiefel} manifolds \cite{james1976topology}, spaces of matrices with orthonormal columns. The tangent space of these manifolds enable a natural connection to the mathematics of ARMA on Lie algebras (i.e., the tangent space to a Lie group) formally developed in \cite{fiori2014auto} and to the Lindblad master equation as shown in App. \ref{app:master}.
Using these relationships, we build a theory of ARMA relevant for quantum information processing that models temporally-correlated noise in a circuit-model formalism. 



%
We begin by showing how CPTP maps are related to Stiefel manifolds. A CPTP map $\Phi$ operating on $N\times N$ density operators $\rho$ can be expressed by
%
%
    $\Phi(\rho) = \sum_{k}M_k\rho M_k^\dagger$
%
%
where $M_k$ are $N\times N$ complex matrices called Kraus operators~\cite{choi1975completely} with $\sum_k M_k^\dag M_k = I_N$.  This form is not unique, but any CPTP map can be expressed using no more than $N^2$ Kraus operators \cite{choi1975completely}. Given $K$ Kraus operators $\{M_k\}$, we define $\mathcal{S}=\begin{bmatrix}M_1^\dagger,\cdots, M_K^\dagger\end{bmatrix}^\dagger$, a $KN\times N$ matrix with orthonormal columns.


Complex $n\times m$ matrices ($n\geq m$) with orthonormal columns define a Stiefel manifold \cite{james1976topology}, denoted $V_{m}(\mathbb{C}^n)$.
Complex Stiefel manifolds are homogeneous spaces of unitary matrix groups, $V_{m}(\mathbb{C}^n)\cong U(n)/U(n-m)$, where $U(n)$ denotes the $n\times n$ unitary matrix (Lie) group. 
Elements of 
$V_{m}(\mathbb{C}^n)$ are associated with equivalence classes of unitary matrices with the first $m$ columns identical. We note there is a known correspondence between Kraus operators and columns of unitary matrices \cite{bruzda2009random,schultz2019exponential, oza2009optimization,pechen2014incoherent,schultz2019exponential}.
Thus, the $KN\times N$ matrix $\mathcal{S} \in V_N(\mathbb{C}^{KN})$ relates Kraus operators to points on the corresponding Stiefel manifold. In order to describe perturbations around this point on the manifold, we introduce the tangent space of $V_N(\mathbb{C}^{KN})$. The tangent vectors are given by matrices 
%
    %
    $X = \mathcal{S}A+\mathcal{S}_\perp B$
    %
%
with with $A$ $N\times N$ skew-Hermitian, $B$ $(KN-N)\times N$ arbitrary,  and $\mathcal{S}_\perp$ denotes an orthogonal complement to $\mathcal{S}$ such that $[\mathcal{S}\,\, \mathcal{S}_\perp]$ is unitary \cite{edelman1998geometry}. 
An exponential map at $\mathcal{S}$ maps $X$ back to the manifold via
\begin{equation}{\label{eq:stief_exp}}
    \exp_\mathcal{S}(X) =
    \begin{bmatrix}\mathcal{S}
    &\mathcal{S}_\perp\end{bmatrix}\exp\left(\begin{bmatrix}A&-B^\dagger\\
    B&\mathbf{0}\end{bmatrix}\right)I_{KN,N}
\end{equation}
where $I_{n,m}$ denotes the first $m$ columns of the $n\times n$ identity matrix $I_n$ , and $\mathbf{0}$ is a zero-matrix of appropriate dimension \cite[Eq.~(2.42)]{edelman1998geometry}.  Therefore, this map provides a mechanism  to generate new elements of $V_N(\mathbb{C}^{KN})$ for specific choices of matrices $A$, $B$ and $\mathcal{S}_\perp$.

For quantum information applications, we typically describe a noisy quantum channel as a perturbation around some desired unitary operation $U$. This corresponds to choosing $\mathcal{S}=[U^\dagger, \mathbf{0}]^\dagger$, and provides a natural choice for $\mathcal{S}_\perp=[\mathbf{0},I_{KN-N}]^\dagger$. 
The exponential map in Eq.~(\ref{eq:stief_exp}) generates the perturbed quantum channel. This form of the map highlights a natural geometric interpretation: $B$ is a block matrix of $K-1$ individual $N\times N$ matrices $B_k$, so $r$ nonzero blocks result in CPTP maps with Kraus rank $K=1+r$, except periodically at the initial point $U$. This connects to the Linblad master equation through the infinitesimal limit of Eq.~(\ref{eq:stief_exp}) that yields the time-evolution of the system density matrix (see App. \ref{app:master} for more details),
\begin{equation}\label{eq:lindblad_schwarma}
\dot{\rho}=-i[H,\rho]+\sum_{k=1}^K\left(B_k\rho B_k^\dagger-\frac{1}{2}\{B_k^\dagger B_k,\rho\}\right),
\end{equation}
where $H=-A/i$.

We now generalize ARMA techniques to the evolution of CPTP maps. Consider a combination of $L$ independent ARMA models with an upper index $\ell$ where $y^{(\ell)}_k$ denotes the output of the $\ell^{\textnormal{th}}$ model at time step $k$, and so on for $a^{(\ell)}_i$, $b^{(\ell)}_j$, and $x^{(\ell)}_k$.  As in standard ARMA, $x_k^{(\ell)}$ are independent zero-mean Gaussian random variables with unit variance (other distributions may be used, as was done in \cite{clader2021impact}).  Let $X^{(\ell)}$ denote fixed elements of the tangent space 
of $\mathcal{U}_k$ in $V_N(\mathbb{C}^{KN})$ corresponding to the target unitary $U_k$.  Then, we define the CPTP map output $\mathcal{S}_k$ of a SchWARMA model at time $k$ by direct generalization of Eq.~(2.20) in \cite{fiori2014auto} from Lie Groups to Stiefel manifolds
\begin{subequations}\label{eq:schwarma}
\begin{align}
\label{eq:schwarmaa}y^{(\ell)}_k &= \sum_{i=1}^{p^{(\ell)}}a^{(\ell)}_iy^{(\ell)}_{k-i} + \sum_{j=0}^{q^{(\ell)}}b_j^{(\ell)} x^{(\ell)}_{k-j} \\
\label{eq:schwarmab}\mathcal{S}_k&=\exp_{\mathcal{U}_k}\left(\sum_{\ell=1}^{L}y_k^{(\ell)} X^{(\ell)}\right)\,,
\end{align}
\end{subequations}
where
\begin{equation}
    \exp_{\mathcal{U}_k}\left(\begin{bmatrix} A\\B\end{bmatrix}\right)=\begin{bmatrix}U&\mathbf{0}\\\mathbf{0}&I_{KN-N}\end{bmatrix}\exp\left(\begin{bmatrix}A& -B^\dagger\\B & \mathbf{0}\end{bmatrix}\right)I_{KN,N}\,,
\end{equation}
for ideal unitary $U$, its corresponding Stiefel representation $\mathcal{U}_k$, and tangent space element 
\begin{equation}
    X=\begin{bmatrix}A\\B\end{bmatrix}\,,
\end{equation}
with $A$ an $N\times N$ skew-symmetric complex matrix, and $B$ an $KN\times N$ arbitrary complex matrix. The $\mathcal{S}_k$ can then be composed as CPTP maps to implement correlated noisy quantum dynamics.

The SchWARMA model provides a coarse-grained representation of time-correlations in a quantum channel.  Specifically, it describes noisy quantum dynamics at discrete time steps (indexed by $k$ in Eq.~\eqref{eq:schwarma} and subsequent examples), such as after circuit gates. To our knowledge, the only other way to do this is via Suzuki-Trotter methods \cite{Suzuki:1976aa, Suzuki:1977aa, Suzuki:1985aa, SUZUKI1990319} to numerically solve the Liouiville equation. This requires step sizes much smaller than the fluctuation time-scale of the Hamiltonian such that $\Delta t \ll ||\partial H(t)/\partial t||^{-1}$ \cite{berry2007}. For high-bandwidth noise or control pulses, this can require many time steps to accurately simulate the trajectory. SchWARMA, on the other hand, estimates the impact of time-dependent noise as an error channel after a perfect gate, with simulation error proportional to the infidelity squared when the noise does not commute with the desired gate. 

By separating control and noise, even in the presence of time-correlations, SchWARMA can provide significant advantage for simulating noisy quantum systems with minimal impact to simulation accuracy. In the following, we discuss examples using SchWARMA models for dephasing and amplitude damping channels, depolarizing channel, as well as continuously driven quantum systems. Details of the error analysis are provided in App. \ref{app:error}. Furthermore, we note that this model-based formulation also offers theoretical paths to mitigate time-correlated noise.

\section{Specific SchWARMA Models} \label{sec:schwarma-models}
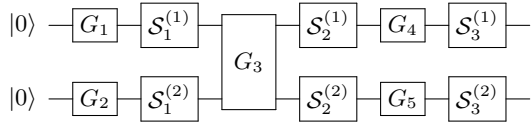
\begin{figure}[h!]
\[
\begin{tabular}{c}\Qcircuit @C=1em @R=1.23em {
& \lstick{\ket{0}} & \gate{G_1} & \gate{\mathcal{S}_{1}^{(1)}} & \multigate{1}{G_3} & \gate{\mathcal{S}_{2}^{(1)}} & \gate{G_4} & \gate{\mathcal{S}_{3}^{(1)}} & \qw \\
& \lstick{\ket{0}} & \gate{G_2} & \gate{\mathcal{S}_{1}^{(2)}} & \ghost{G_3}        & \gate{\mathcal{S}_{2}^{(2)}} & \gate{G_5} & \gate{\mathcal{S}_{3}^{(2)}} & \qw \\
} \end{tabular}
\]
\caption{Circuit schematic demonstrating how one might implement a SchWARMA model within a circuit of arbitrary one and two qubit gates labelled with $G_i$. It allows one to consider perfect gates followed by noise that retains the time correlation. The $S_k^{(q)}$ terms are the error channel defined in Eq.~\eqref{eq:schwarmab} with additional upper index denoting the qubit.} 
\label{fig:schwarma_circuit}
 
\end{figure}

Here we provide details of how we implement the SchWARMA models for the various examples we provide in the next section. First, consider an example where we wish to simulate correlated noise within a quantum circuit such as the example shown in Fig.~\ref{fig:schwarma_circuit} that contains arbitrary one and two qubit gates specified as $G_i$. For the noise model, we create two SchWARMA models (denoted by an additional upper index on $\mathcal{S}_k$) given by Eq.~\eqref{eq:schwarmab} with $U=I_{2}$ chosen to be the identity as we want to sample our error channel near the identity.  This allows one to simply insert random noise channels into the circuit with the amplitude of the noise determined by the ARMA coefficients given in Eq.~\eqref{eq:schwarmaa}. Specific choices of noise channel are considered next.

\subsection{Z Dephasing}\label{sec:single}
We define a $Z$ dephasing SchWARMA model with:
\begin{subequations}\label{eq:zdephasing}
\begin{align}
U & = \begin{bmatrix}
    1 & 0 \\
    0 & 1
\end{bmatrix}, \\
X = -i \sigma_z & = \begin{bmatrix}
    -i & 0 \\
    0 & i
\end{bmatrix},
\end{align}
\end{subequations}
so $N=2$ and $K=1$, corresponding to unitary dynamics on a qubit.  Since $K=1$, we have that $\exp_{\mathcal{U}_k}$ is the usual matrix exponential and
this gives the map to apply at circuit location $k$ by the simple expression
\begin{equation}
    \label{eq:schwarma_dephasing}
    \mathcal{S}_k = \begin{bmatrix}
        e^{-i y_k} & 0 \\
        0 & e^{i y_k}
    \end{bmatrix},
\end{equation}
where we have suppressed the $\ell$ index, since there is only a single SchWARMA model needed. This is a $Z$ rotation by a random angle given by a classical ARMA model.

\subsection{Multi-axis Dephasing}\label{sec:multi}
For the slightly more complicated multi-axis noise that we consider for our surface code simulations in the next section, the SchWARMA model is given by:
\begin{subequations}\label{eq:xyzdephasing}
\begin{align}
U & = \begin{bmatrix}
    1 & 0 \\
    0 & 1
\end{bmatrix}, \\
X^{(x)} = -i \sigma_x & = \begin{bmatrix}
    0 & -i \\
    -i & 0
\end{bmatrix},\\
X^{(y)} = -i \sigma_y & = \begin{bmatrix}
    0 & -1 \\
    1 & 0
\end{bmatrix},\\
X^{(z)} = -i \sigma_z & = \begin{bmatrix}
    -i & 0 \\
    0 & i
\end{bmatrix}\,.
\end{align}
\end{subequations}
Again, $N=2$ and $K=1$, corresponding to unitary dynamics, but this time with $L=3$ corresponding to multiple underlying ARMA models.  Again,  we have that $\exp_{\mathcal{U}_k}$ is the usual matrix exponential for qubit unitary dynamics, and this gives the map to apply at circuit location $k$ by the expression
\begin{align}
    \label{eq:schwarma_xyzdephasing}
    \mathcal{S}_k & = \exp[-i (y_k^{(x)} \sigma_x + y_k^{(y)} \sigma_y + y_k^{(z)} \sigma_z)] \\
     & = \begin{bmatrix}
        \cos(g_k) - i y_k^{(z)} \textnormal{sinc}(g_k) & -(i y_k^{(x)} - y_k^{(y)}) \textnormal{sinc}(g_k) \\
        -(i y_k^{(x)} + y_k^{(y)}) \textnormal{sinc}(g_k) & \cos(g_k) + i y_k^{(z)} \textnormal{sinc}(g_k) 
    \end{bmatrix},
\end{align}
where $g_k = \sqrt{y_k^{(x)^2} + y_k^{(y)^2} + y_k^{(z)^2}}$, $N=1$, and $\textnormal{sinc}(x) = \sin(x)/x$.

\subsection{Amplitude Damping}
Finally, we consider the SchWARMA model for the non-unitary amplitude damping channel. The tangent space element for that SchWARMA model is :
\begin{equation}\label{eq:app:damp_tangent}
    \begin{aligned}
            X&=\begin{bmatrix}
                \multicolumn{2}{l}{\multirow{2}{*}{$\mathbf{0}_{2\times 2}$}}\\[.5em]
                0 & 1 \\
                0 & 0 \end{bmatrix}
    \end{aligned}
\end{equation}
where the notation $\mathbf{0}_{m\times n}$ indicates a matrix of zeros with $m$ rows and $n$ columns, $N=2$, and $K=2$. Thus $S_k$ is given by:
\begin{equation}
    \label{eq:app:ad_dephasing}
    \mathcal{S}_k = \exp\left(y_k \begin{bmatrix}
                \multicolumn{2}{c}{\multirow{2}{*}{$\mathbf{0}_{2\times 2}$}} & 0
                &0\\
                && -1 & 0\\
                0 & 1 & \multicolumn{2}{c}{\multirow{2}{*}{$\mathbf{0}_{2\times 2}$}}\\
                0 & 0 \\
            \end{bmatrix} \right) I_{4,2},
\end{equation}
yielding an $4 \times 2$ sized matrix containing the two $2 \times 2$ Kraus operators.  Since the tangent space element $X$ in \eqref{eq:app:damp_tangent} has all zeros in the skew-symmetric $A$ component, in this case the driving ARMA model $y_k$ can actually be complex-valued.  These two Kraus operators at SchWARMA index $k$ are:
\begin{subequations}
\begin{align}
    M_1^{(k)} & = \begin{bmatrix}
    1 & 0 \\
    0 & \cos (|y_k|)
    \end{bmatrix}, \\
    M_2^{(k)} & = \begin{bmatrix}
    0 & \sin (|y_k|)e^{i\angle y_k} \\
    0 & 0
    \end{bmatrix}\,. \\ \nonumber
\end{align}
\end{subequations}

We see that this SchWARMA model is nothing more than the standard amplitude damping model, but with a decay coefficient given by the square of the sin of the absolute value of the classical ARMA model coefficient. This SchWARMA model simulates an amplitude damping channel with a correlated, fluctuating decay rate, allowing for the analysis of temporal fluctuations in the decay rate of the qubit \cite{klimov2018fluctuations}. However, it does not consider time-dynamics of the system-bath interaction which is still assumed to happen on infinitely fast time-scales. 

We note that this process allows for some quantum processes that appear to be non-physical with, for example, negative amplitude damping coefficients. All channels created using this method are CPTP and physical. Since manifolds are only locally linear, when one takes a large step in the tangent space, the curvature of the manifold is not respected. Therefore, a direction that appears to be amplitude damping at the start of the curve will appear like something else relative to some point further along the curve defined by the arc.

Therefore, for $|y_k| < \pi/2$ we have classic amplitude damping, which includes both the non-unital shift towards $|0\rangle$ as well as anisotropic contractions in the $X$, $Y$, and $Z$ dimensions of the Bloch sphere to maintain physicality. For $y_k = \pm\pi/2$ all states get mapped to the $\ket{0}$ state. For $y_k$ between $\pi/2$ and $\pi$, the non-unital shift shrinks back to the origin and the contraction process reverses, except that the $X$ and $Y$ axes are both flipped. At $y_k = \pi$ we obtain the $Z$ gate. This sequence is then reversed for $\pi < y_k < 2\pi$. The primary point here is that for non-unitary channels such as the amplitude damping channel, the step size should be small enough such that $0 \leq |y_k| < \pi/2$. For high-fidelity quantum operations, this will never be a concern.

\subsection{Lindbladian Dephasing and Depolarizing}
When averaged over random instances, the stochastic unitary SchWARMA models in sections \ref{sec:single} and \ref{sec:multi} above will produce, at each time step $k$, dephasing channels
\begin{equation}
    \rho\to (1-p_z)\rho+p_z\sigma_z\rho \sigma_z^\dagger
\end{equation}
and depolarizing channels
\begin{equation}
    \rho \to (1-p_x-p_y-p_z)\rho+\sum_{j=x,y,z}p_j\sigma_j\rho\sigma_j^\dagger\,,
\end{equation}
respectively, where the $p_j$ are determined by the statistics of the
underlying ARMA models at step $k$.  At the SchWARMA level, this is
fundamentally different from Lindblad style decay into an infinitely fast bath
using Lindblad operators as described in App. \ref{app:master} (i.e., $B_k\neq 0$).  This is evident, for example, in the fact that SchWARMA errors corresponding to stochastic unitary evolution offer the potential of control (for time-correlated errors), whereas SchWARMA errors motivated by Lindblad evolutions will be purely dissapative regardless of the spectrum of the driving SchWARMA model.  Furthermore, the correspondence between SchWARMA tangent vectors and the Lindblad master equation tells us that these Lindbladian-style depolarizing channels can be simulated in SchWARMA by $U=I_2$ and
\begin{subequations}\label{eq:lind_x}
    \begin{align}
            X^{(x)}&=\begin{bmatrix}
                \mathbf{0}_{2\times 2}\\
                \sigma_x\\
                \mathbf{0}_{4\times 2}\end{bmatrix}\,,\\
            X^{(y)}&=\begin{bmatrix}
                \mathbf{0}_{4\times 2}\\
                \sigma_y\\
                \mathbf{0}_{2\times 2}\end{bmatrix}\,,\\
            X^{(z)}&=\begin{bmatrix}
                \mathbf{0}_{6\times 2}\\
                \sigma_z\end{bmatrix}\,,
    \end{align}
\end{subequations}
with corresponding (possibly complex-valued) ARMA outputs $y_{k}^{(j)}$ for $j=x,y,z$.   This model results in SchWARMA steps
\begin{equation}
    \mathcal{S}_k=\exp\left(X_k^{(ES)}\right)I_{8,2}\,,
\end{equation}
where
\begin{equation}
    X_{k}^{(ES)} = \begin{bmatrix}
    \mathbf{0}_{2\times 2} & -(y_k^{(x)}\sigma_x)^\dagger & -(y_k^{(y)}\sigma_y)^\dagger & -(y_k^{(z)}\sigma_z)^\dagger\\
    y_k^{(x)}\sigma_x & \multicolumn{3}{c}{\multirow{3}{*}{$\mathbf{0}_{6\times 6}$}}\\
    y_k^{(y)}\sigma_y\\
    y_k^{(z)}\sigma_z\\
    \end{bmatrix}
\end{equation}
(c.f., Eq.~\ref{eq:full_unitary}) and the resulting error can be verified numerically to be four Kraus operators corresponding to a depolarizing channel. Clearly, the $x$ and $y$ terms can be dropped off and the $z$ element ``compacted'' to a $4\times 2$ matrix to restrict the errors to dephasing.

\section{Examples}

We now  demonstrate the utility of SchWARMA for analyzing correlated noise within a variety of domains relevant to quantum information processing.  We first build from single-qubit examples in quantum noise spectroscopy and control, showing strong agreement with established theory. Then we show the utility of SchWARMA in a multi-qubit circuit simulation, as it results in substantial computational savings over a stochastic Liouville approach. Code to reproduce these examples and others can be found in our \texttt{mezze} package \cite{mezze}.


\subsection{Example 1: Quantum Noise Spectroscopy}
In this first example, we verify the SchWARMA noise spectrum is consistent with quantum noise spectroscopy. 
%
%
%
We model the dephasing dynamics of a qubit with the stochastic Hamiltonian $H(t)=\eta(t)\sigma_z+\Omega(t)\sigma_x$ using SchWARMA, where $\eta$ is a stationary and Gaussian semi-classical noise and $\Omega$ infinitely fast control (as in a quantum circuit model). Stochastic dephasing noise leads to decay in the survival probability $p$ of a qubit both prepared and measured in the $|+\rangle$ state (assuming no state preparation and measurement error), given by $p=1/2[1+\exp(-\int d\omega |F(\omega)|^2 S_y(\omega))]$
%
    %
    %
%
where $F(\omega)$ is the filter function (Fourier transform of the control's modulation function, for single axis $X$ control) \cite{cywinski2008:fff, PhysRevLett.113.250501}.  Under these assumptions, the noise spectrum can be inferred by applying different control inputs. Here, we use different gate sequences to define a regression problem in the fashion of \cite{alvarez2011:qns, szakowski2017:qns, pazsilva2017:qns}.

To perform effective quantum noise spectroscopy, it is necessary to have spectrally diverse (in terms of the filter function) control sequences. To generate these, we used $W/2$ different sequences of common length $W$ gates (128 for the bandlimited and multipole noise, 2048 for the $f^{-\alpha}$ noise) whose discrete-time modulation functions \cite{cywinski2008:fff, PhysRevLett.113.250501} were defined by $f_k(m)=\textrm{sign}(\cos(\pi (k-1)m/W))$ for $k,m=1,\dots,W/2$, meaning that an $X$ gate is applied when the sign of the modulation function changed, or an $I$ gate otherwise.  Assuming ideal state preparation and measurement in $|+\rangle$, the Fourier transform of $f_k(m)$, $F_k(\omega)$, can be used to define a system of equations to estimate the power spectrum $S_Y(\omega)$ by inverting the system of equations $S_Y(2k\pi/W)=\sum_m |F_m(2k\pi/W)|^2\chi_{m}$, where $\chi_m -\log(2\hat{p}_m-1)$, with each control sequence repeated sufficiently many times to achieve an accurate estimate of the survival probability $p_m$.  Note that nonnegative least squares should be used to ensure nonnegativity of the estimated power spectrum. Additionally, in order to reduce sampling error, we did not perform repeated Bernoulli trials to construct $\hat{p}_m$, instead we took the average of the exact survival probabilities over 1000 independent SchWARMA trajectories.

\begin{figure}
\centering
\includegraphics[height=3.36cm]{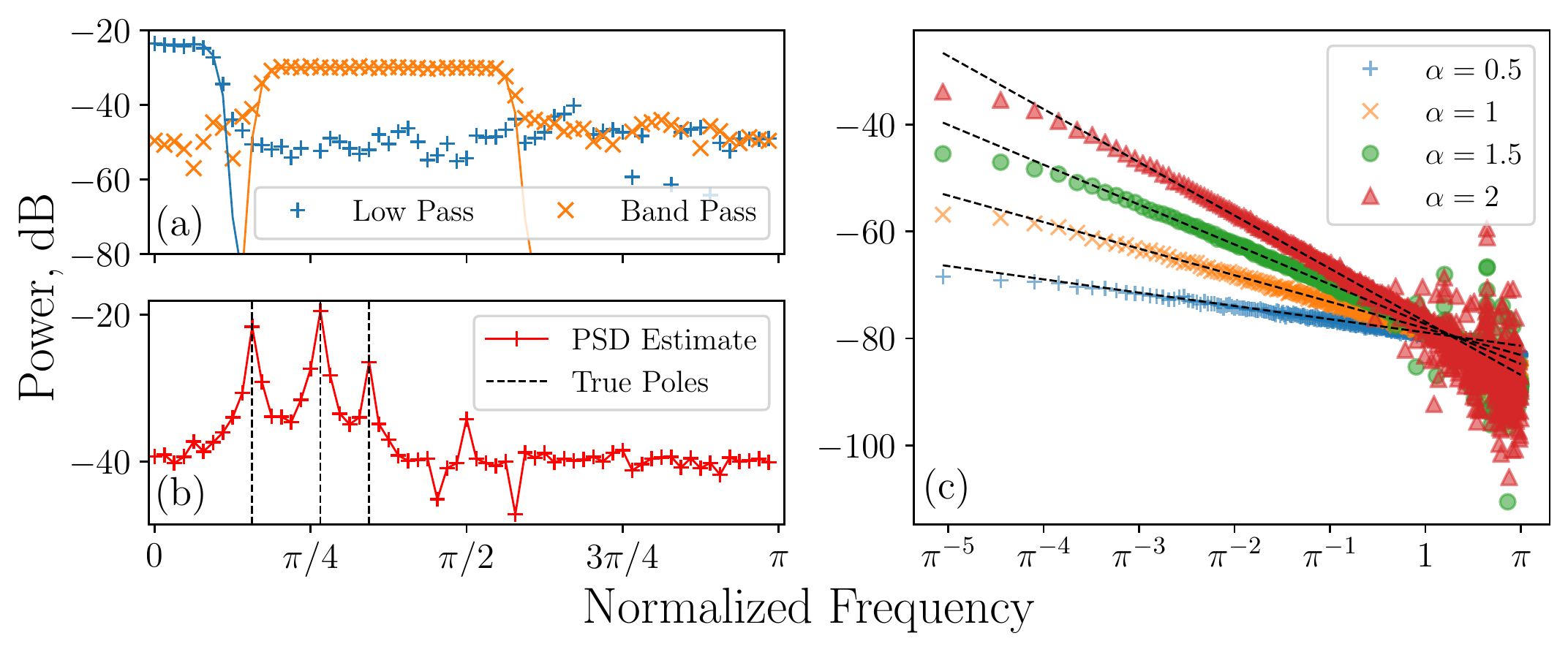}

\caption{Simulated QNS experiments. (a) Spectrum reconstructions (markers) of SchWARMA simulated low-pass and band-pass noise (lines).  (b) Spectrum reconstruction of SchWARMA simulated ``peaky'' multi-pole AR spectrum.
(c) Spectrum reconstructions of SchWARMA simulated $f^{-\alpha}$ noise, dashed lines indicate ideal $\alpha$.}

\label{fig:psds}

\end{figure}

Exploiting the universality of ARMA to produce \textit{any} power spectra, SchWARMA simulations of quantum noise spectroscopy for several spectra are shown in Fig.~\ref{fig:psds}.  The noise generation was handled using $Z$-dephasing SchWARMA models.  The bandlimited signals were MA models defined by a 128-tap Hamming window of desired bandwidth and cosine modulated to the appropriate center frequency (as needed).  The multipole noise was generated by a purely AR SchWARMA model with coefficients set by placing three complex poles of the transfer function on the unit circle  at locations corresponding to the desired peaks in the normalized frequency space and their complex conjugates.  SchWARMA coefficients for the $f^{-\alpha}$ noise were generated using the methods in \cite{plaszczynski2007generating}. The resulting estimates strongly agree with the true spectra, with the primary limiting factors being Monte Carlo accuracy and conditioning of the regression equations.

\subsection{Example 2: Dynamical Decoupling}
Next, consider a single qubit in the presence of multi-axis noise and control,
\begin{equation}
    \label{eq:circuit_hamiltonian}
    H(t) = \sum_{i=\{x,y,z\}}\eta_i(t) \sigma_i + \Omega(t)(\cos(\phi)\sigma_x + \sin(\phi)\sigma_y),
\end{equation}
where $\eta_i(t)$ are zero-mean wide-sense stationary Gaussian processes with correlation functions $\langle \eta_i(\tau)\eta_j(t)\rangle = f_i(|\tau-t|)\delta_{ij}$, $\phi$ is a user-controlled parameter that specifies the axis of rotation, and $\Omega(t)$ is the time-dependent amplitude of the control pulse. 
Furthermore, let the $\eta_i$ have correlation functions  
\begin{equation} \label{eq:gaussian_ac}
f_i(\tau) =  \frac{\gamma_i}{2\sqrt{\pi}\tau_{c,i}}\exp\left(\frac{-\tau^2}{4\tau_{c,i}^2}\right), 
\end{equation}
where $\tau_{c,i}$ is the correlation-time of the noise for axis $i$ and $\gamma_i$ is the corresponding noise-amplitude.

\begin{figure}
\centering
\includegraphics[width=8.4cm,height=3.36cm]{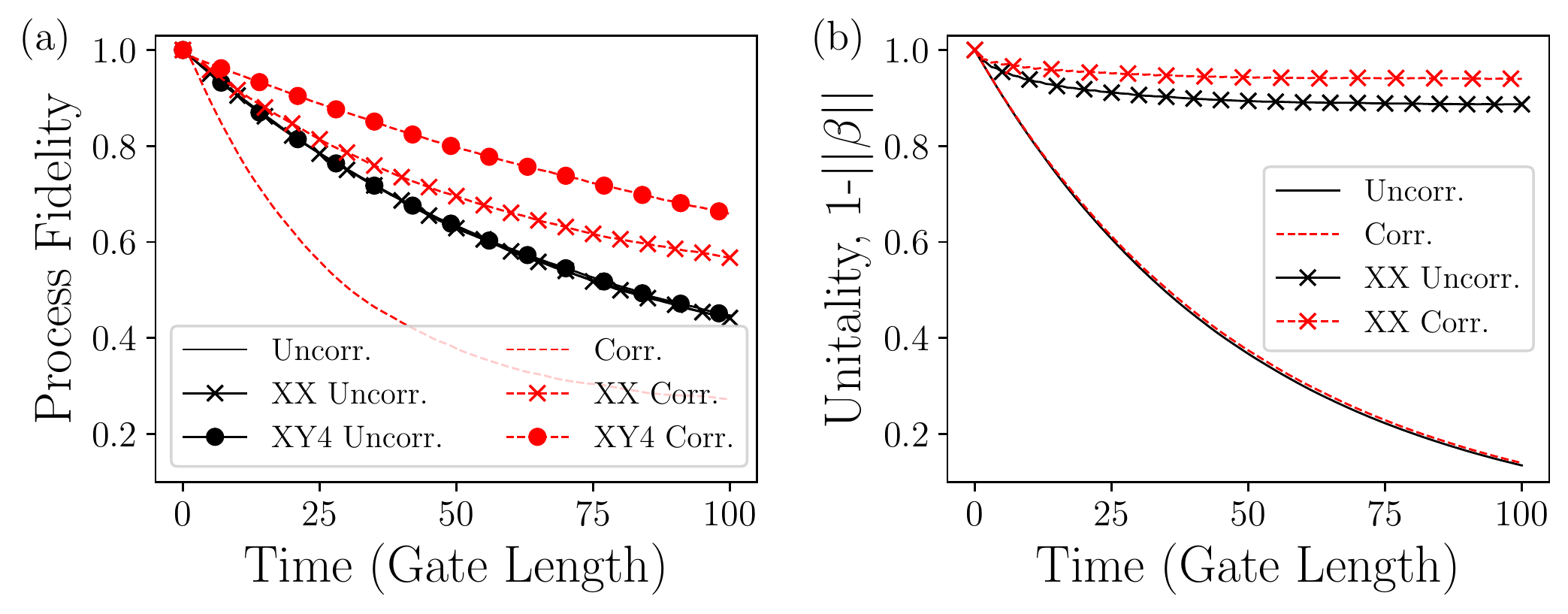}

\caption{SchWARMA control simulations. (a) Process fidelity decay due to multi-axis coherent noise. Correlated noise $(\tau_{c}=3)$ produces more rapid decay than uncorrelated noise. Decoupling protocols reduce the decay for correlated case, with slower decay for $XY4$ than $XX$. 
(b) Decay in unitality from amplitude damping.  Uncontrolled decay is essentially identical, but decoupling protocols better preserve unitality for correlated damping ($\tau_{c}=3$).  In this case, $XX$ and $XY4$ (not shown) are identical.}
\label{fig:ctrl}
\end{figure}

Decoupling  control  protocols  can  be  an  effective  mechanism for reducing decoherence. The key to their error suppression is the fact that no error  process happens instantaneously, despite this being a common approximation in open quantum systems theory. If one can control the system on a time-scale fast relative to the error dynamics, then error suppression is possible with these protocols \cite{viola1998:dd, viola1999:dd, zanardi1999:dd, byrd2002:dd}.
Furthermore, some protocols are more effective at suppressing different classes of noise compared to others.  In particular, the $XY4$ protocol \cite{maudsley1986:xy4} suppresses multi-axis semi-classical errors to second order, whereas repeated $XX$ gates only suppress $Y$ and $Z$ errors to second order.  Fig.~\ref{fig:ctrl} (left) shows SchWARMA reproducing these phenomena in terms of process fidelity, using the Hamiltonian in Eq.~\ref{eq:circuit_hamiltonian} with separate ARMA models driving the $\eta_i$, and instantaneous $\pi$ control pulses. The process fidelity $F_p(U,\Phi)$ at each time-step between the ideal unitary operator $U$, and the average noisy map, is defined as $(1/N^2) Tr [(U^\top \otimes U^\dagger)(\sum_k M_k^* \otimes M_k)]$, where $*$ and $\top$ denote complex conjugation and transpose, respectively.

For amplitude damping, decoupling protocols will not reduce decoherence here as we assume an infinitely fast bath, but consider the \textit{non-unital} action of the map.  All CPTP maps have an affine form  \cite{fujiwara1999one}, visualized in Bloch space as a combination of a unital contraction and rotation operations 
along with a non-unital shift from the origin. This shift is defined by the vector $\beta$, $\beta_i = \langle\langle I_N|\sum M_k^* \otimes M_k|\sigma_i\rangle\rangle$, where $|\cdot\rangle\rangle$ denotes column stacking and $\langle\langle \cdot|=|\cdot\rangle\rangle^\dagger$. For amplitude damping $\beta$ shifts the origin towards $|0\rangle$. However, decoupling sequences will flip the direction of this shift, resulting in slower rate of unitality (1-$||\beta||$) decay, as depicted in Fig.~\ref{fig:ctrl} (right). This demonstrates SchWARMA simulating the impact of temporally-correlated classical noise processes that affect the rate of decoherence even for non-unital channels such as those observed in \cite{klimov2018fluctuations}.

\subsection{Example 3: Quantum Circuit Simulation}
%

 Next, to demonstrate the utility of using SchWARMA to simulate noisy multi-qubit quantum circuits, we model $Z$ and $X$ check circuits for the surface code \cite{PhysRevA.86.032324}. We compare an approach using both a full density-matrix simulation of the noisy pulse sequence with a first-order Trotter-Suzuki technique \cite{Suzuki:1976aa, Suzuki:1977aa, Suzuki:1985aa, SUZUKI1990319}, and a SchWARMA simulation. The SchWARMA simulation assumes the gates to be instantaneous and follows each gate with a random unitary rotation from the SchWARMA model: $\mathcal{S}_k^{(j)}=\exp(-i\sum_{\mu=x,y,z} y_k^{(j,\mu)}\sigma^{(j)}_\mu)$ for each qubit $j$ at time-step $k$.  
 
For the full Trotter-based simulation, each single-qubit gate is generated by a Hamiltonian:
\begin{equation}
    \label{eq:circuit_hamiltonianSUP}
    H(t) = \sum_{i=\{x,y,z\}}\eta_i(t) \sigma_i + \Omega(t)(\cos(\phi)\sigma_x + \sin(\phi)\sigma_y),
\end{equation}
where $\eta_i(t)$ are zero-mean wide-sense stationary Gaussian processes with correlation functions $\langle \eta_i(\tau)\eta_j(t)\rangle = f_i(|\tau-t|)\delta_{ij}$, $\phi$ is a user-controlled parameter that specifies the axis of rotation, $\Omega(t)$ is the time-dependent amplitude of the control pulse, and $\eta_i$ is defined in Eq. \eqref{eq:gaussian_ac}. Entanglement in the system is generated via a two qubit $ZZ$ entangling operation
\begin{equation}
    \label{eq:two_qubit_gate}
    H(t) = \Omega^{(k,j)}_{ZZ}(t) \sigma_z^{(k)}\otimes \sigma_z^{(j)},
\end{equation}
where $\Omega^{(k,j)}_{ZZ}(t)$ is a controllable coupling term and $k$ and $j$ denote the qubit index. To create equivalent noise models between the Trotter simulation and SchWARMA simulation, we used the methods outlined in App. \ref{app:coeffs}.

\begin{figure}
\[
\begin{tabular}{ccc}
\begin{tabular}{c}\Qcircuit @C=1em @R=1.5em {
& \ctrl{1} & \qw \\
& \targ    & \qw
} \end{tabular} & = &
\begin{tabular}{c}\Qcircuit @C=1em @R=.7em {
& \qw             & \qw      & \multigate{1}{ZZ_{90}} & \gate{Z^{1/2}} & \qw      & \qw            & \qw \\
& \gate{Y^{-1/2}} & \gate{X} & \ghost{ZZ_{90}}        & \gate{Z^{1/2}} & \gate{X} & \gate{Y^{1/2}} & \qw 
} \end{tabular}
\end{tabular}
\]
\caption{\label{fig:app:cnot_compilation}The CNOT decomposed into elementary gates. The $Z$ rotations are taken to be virtual within all of our simulations and take no time and occur without error.}
\end{figure}
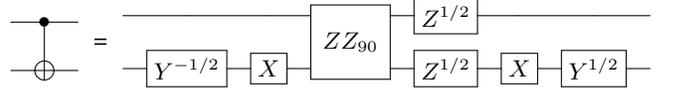

The standard $X$ and $Z$ check circuits for the surface code are given in Fig. \ref{fig:app:surf_checks}:
\begin{figure}
\[
\begin{tabular}{ccc}
\begin{tabular}{c}\Qcircuit @C=1em @R=.7em {
& \lstick{\ket{0}} & \gate{H} & \ctrl{2} & \ctrl{1} & \ctrl{4} & \ctrl{3} & \gate{H} & \qw \\
& \lstick{}        & \qw      & \qw      & \targ    & \qw      & \qw      & \qw      & \qw \\
& \lstick{}        & \qw      & \targ    & \qw      & \qw      & \qw      & \qw      & \qw \\
& \lstick{}        & \qw      & \qw      & \qw      & \qw      & \targ    & \qw      & \qw \\
& \lstick{}        & \qw      & \qw      & \qw      & \targ    & \qw      & \qw      & \qw
} \end{tabular} & \quad &
\begin{tabular}{c}\Qcircuit @C=1em @R=1.23em {
& \lstick{\ket{0}} & \targ     & \targ     & \targ     & \targ     & \qw \\
& \lstick{}        & \qw       & \ctrl{-1} & \qw       & \qw       & \qw \\
& \lstick{}        & \ctrl{-2} & \qw       & \qw       & \qw       & \qw \\
& \lstick{}        & \qw       & \qw       & \qw       & \ctrl{-3} & \qw \\
& \lstick{}        & \qw       & \qw       & \ctrl{-4} & \qw       & \qw
} \end{tabular}
\end{tabular}
\]
\caption{\label{fig:app:surf_checks} The $X$ stabilizer is given on the left and the $Z$ stabilizer is given on the right. Measurements of the ancilla are omitted from these circuits.}
\end{figure}
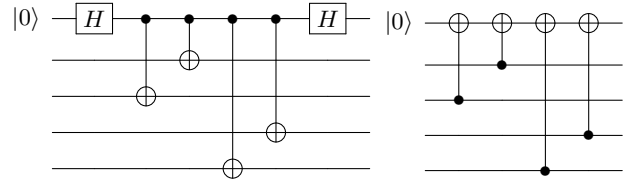
We compile the CNOT into the circuit shown in Fig. \ref{fig:app:cnot_compilation}. The gates are defined as: $$X=e^{-i \sigma_x \pi/2},$$ $$Y^{\pm 1/2} = e^{\mp i \sigma_y \pi/4},$$ $$Z^{1/2} = e^{-i\sigma_z \pi/4},$$ and the controlled $ZZ$ gate between qubits $i$ and $j$ is $$ZZ_{90} = e^{-i \sigma_z^{(i)}\otimes\sigma_z^{(j)} \pi/4}.$$ For the $Z$ check operator, all the single qubit $X$ and $Y$ rotations between the CNOT gates cancel each other. This allows for a further simplification of the compiled circuit of just single-qubit rotations on the ancilla to start, a sequence of $ZZ_{90}$ gates, followed by single qubit rotations on the ancilla. There are leftover virtual $Z^{1/2}$ gates to be done on the data qubits, but we leave these as they would be virtual and taken care of during the next round of error correction in any system. We chose this gate-set as an exemplar and with the knowledge that it is a commonly seen Hamiltonian in experimental systems.

\begin{figure}[h!]
    \centering
    \includegraphics[width=8.4cm]{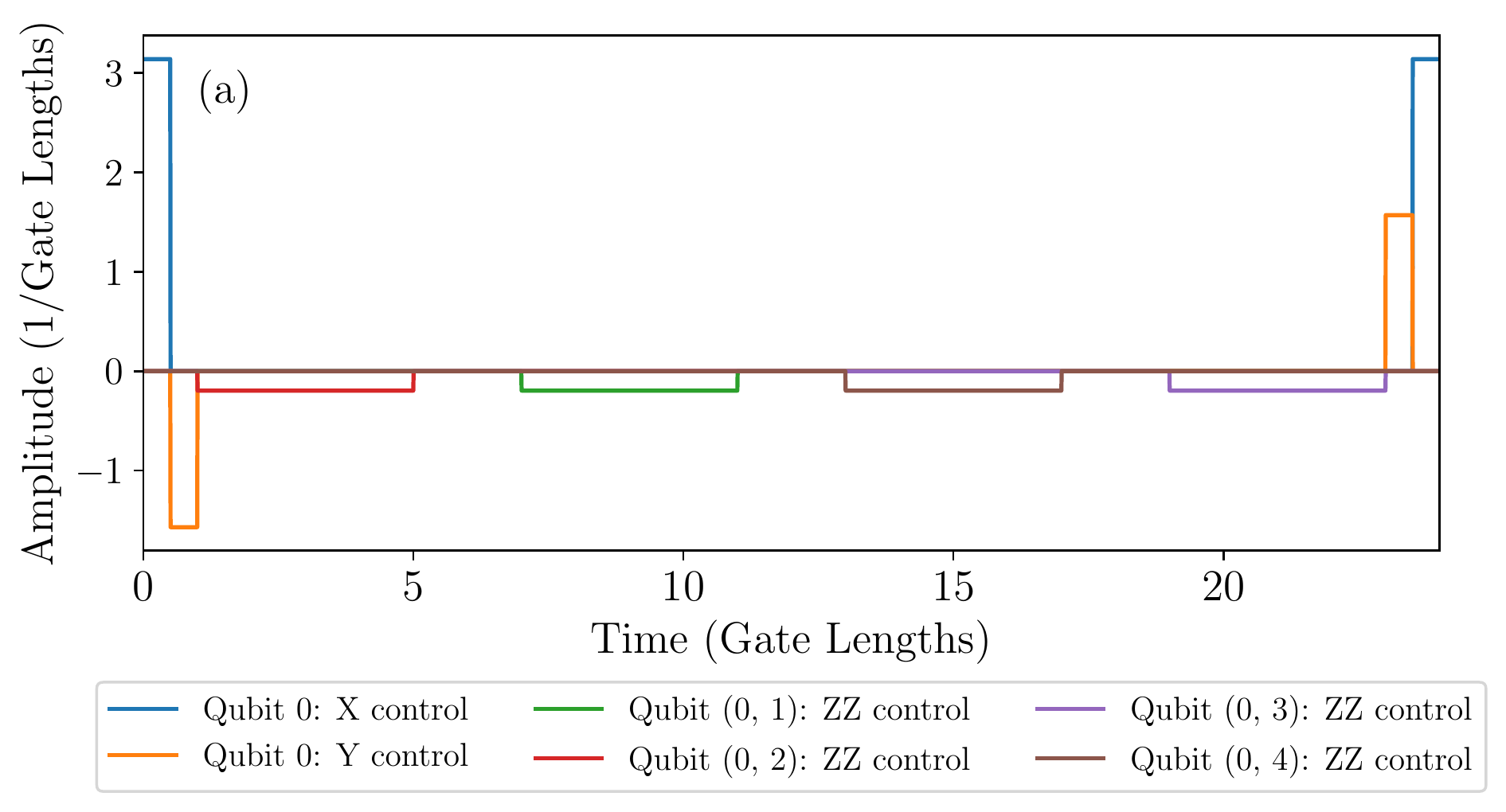}
    \includegraphics[width=8.2cm]{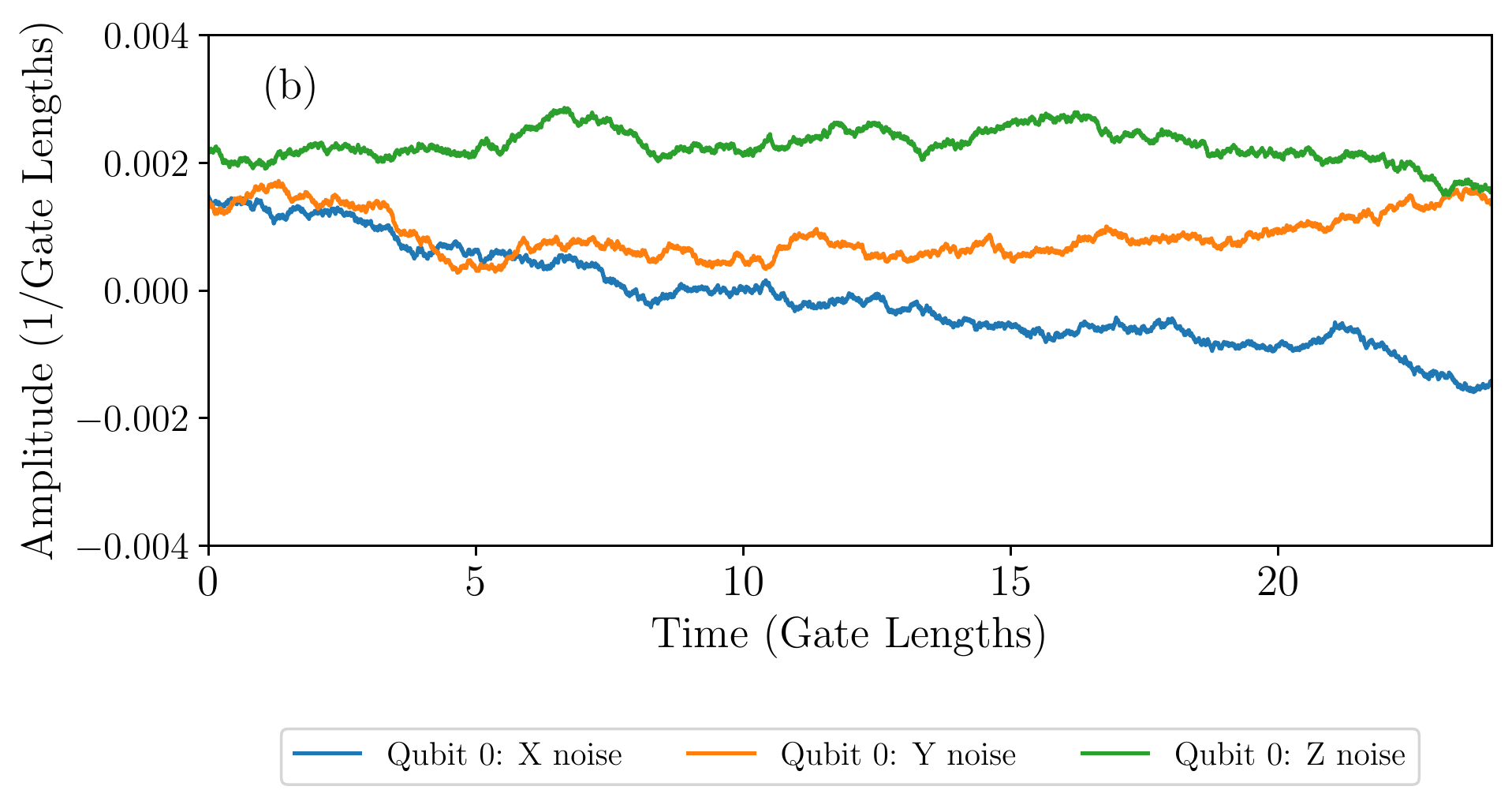}
    \caption{These two plots provide an illustrative example of how the full simulation of the surface code circuits are done. In (a), we show the single-qubit control amplitudes, $\Omega(t)$, for the Hamiltonian in Eq.~\eqref{eq:circuit_hamiltonianSUP} and the two-qubit control amplitudes, $A(t)$, for the Hamiltonian in Eq.~\eqref{eq:two_qubit_gate} for the qubits labelled in the legend. We see single-qubit $X$ and $Y$ controls being applied at the beginning and end of the circuit. These correspond to the application of the Hadamard gate for the $X$ stabilizer. In the middle, the entangling operations are performed. In (b), we plot a single realization of a noise trajectory, $\eta_i(t)$, for the multi-axis noise term in Eq.~\eqref{eq:circuit_hamiltonianSUP} for the same qubit used in (a). We choose noise values $\gamma_x = \gamma_y = \gamma_z = 10^{-4}$ and the correlation time $\tau_{c,x} = \tau_{c,y}=\tau_{c,z} = 32$. Those are in units of inverse gate length and gate length respectively. The noise is always on across the entire duration of the circuit. These values yield a highly time-correlated trajectory.}
    \label{fig:surface_code_noise}
\end{figure}

For the full density matrix simulation, we simulate the entire circuit using a first-order Trotter-Suzuki \cite{Suzuki:1976aa, Suzuki:1977aa, Suzuki:1985aa, SUZUKI1990319} approximation to the dynamics. We generate rectangular pulses for each of the gates in our gate-set. We average over different noise realizations and compute for average process map for the entire circuit. An illustrative example is provided in Fig.~\ref{fig:surface_code_noise} where we plot the single and two-qubit control pulses for qubit 0 (the top qubit in the circuit diagram) and a single realization of a noise trajectory for the terms in Eqs. \eqref{eq:circuit_hamiltonianSUP} and \eqref{eq:two_qubit_gate}. 

In contrast, the SchWARMA simulations are much simpler. For these, we do standard circuit simulation where we simply apply the gates as perfect unitary operations and follow them with SchWARMA generated noise after each gate. We average over the same number of noise realizations as in the Trotter-Suzuki simulation method. This does not require generating the full pulse wave forms, as we had to for the Trotter method, yet provides the same circuit level noise model.

\begin{figure}
\centering
\includegraphics[width=8.4cm, height=3.36cm]{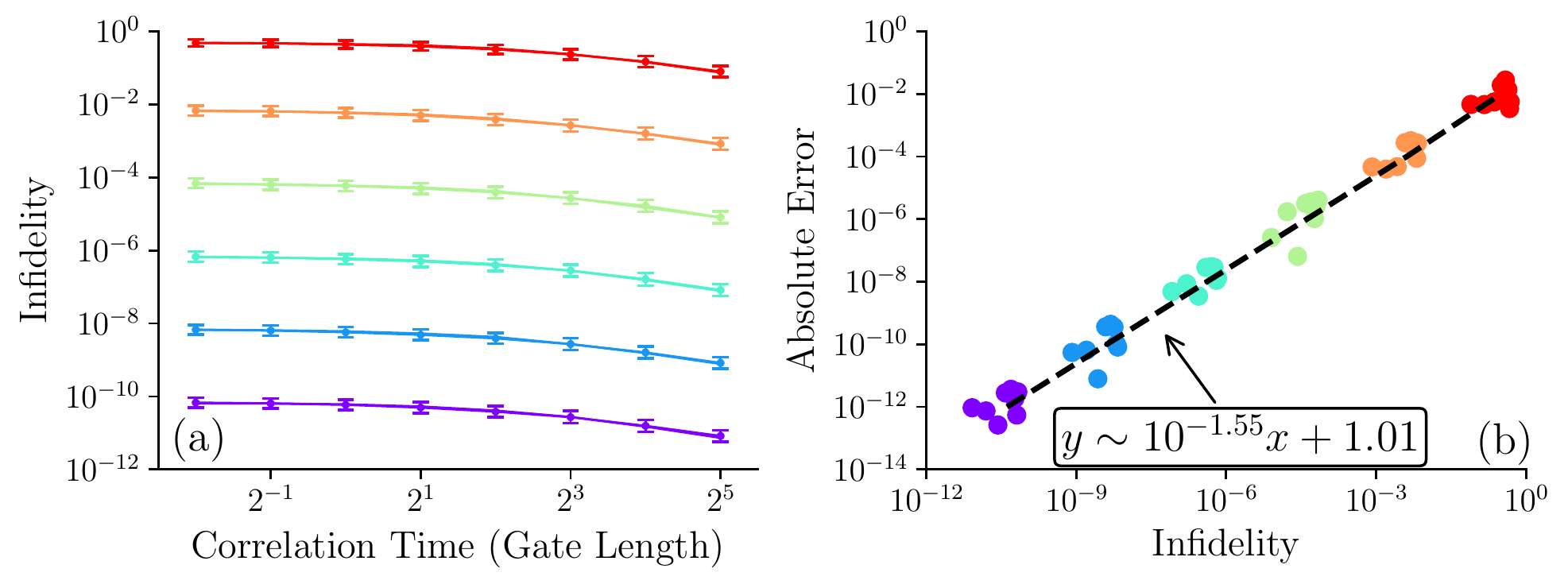}
\caption{Process infidelities for surface code $Z$ check circuits (a) and the absolute error between the SchWARMA and Trotter simulations (b) as a function of noise parameters $\tau_{c,i}$ ((a) $x$-axis), $\gamma_i$ (color) defined below Eq.~\eqref{eq:circuit_hamiltonian}, and process infidelity ((b) $x$-axis). 
The noise strengths are $\gamma_{x,y,z}\in\{10^{-12}, 10^{-10}, \cdots, 10^{-2}\}$ (bottom to top on (a) plot) in units of inverse gate length. In (a), the Trotter simulations are the line plot while the SchWARMA simulations are the marks with Monte Carlo (1000 samples) error bars. In (b), the absolute difference between the SchWARMA and Trotter simulations are plotted with a fit line that is consistent with Monte Carlo sampling error.}
\label{fig:surfplots}
\end{figure}

We compare the process infidelity $1-F_p$ of Monte Carlo averaged CPTP maps for a five qubit circuit and plot the absolute error  (difference in infidelity) between the SchWARMA and Trotter simulation methods in Fig.~\ref{fig:surfplots}  (1000 Monte Carlo trials per point) for  different noise  amplitudes and correlation times for the $Z$ check circuit (The $X$ check circuit results are nearly identical and not displayed). SchWARMA shows very good agreement with the full Trotter simulation. The error observed between the two methods is consistent with Monte Carlo sampling error as the slope of the error fit is $10^{-1.5} \approx \sqrt{1/1000}$. Error proportional to the infidelity squared due to the SchWARMA model is negligible compared to sampling error when the noise is low, see \ref{app:error}. The SchWARMA simulations take $~1/\Delta t$ less computational time, where $\Delta t$ is the Trotter time step, which in our case is a factor of 1000 computational speedup. 
This example highlights the utility of using SchWARMA to simulate correlated noise within quantum circuits, with a complexity no worse than a state-vector simulator such as the ones used in Refs. \cite{PhysRevA.90.062320, PhysRevA.95.062338}.

\subsection{Example 4: Continuously Driven Systems} \label{app:adiabatic}
SchWARMA can also be utilized to simulate continuously driven system dynamics in the presence of correlated noise, such as faulty adiabatic dynamics. We demonstrate this by simulating \begin{equation}
    H(t)=H_{c}(t)+H_{e}(t),
\end{equation}
where $H_{c}(t)$ generates the pure continuously driven (controlled) dynamics and $H_e(t)$ is the error-generating Hamiltonian. Here, $H_e(t)$ describes a temporally correlated noise process. Generically, we can represent the error Hamiltonian as $H_E(t)=\sum_k \eta_k(t) S_k$ with $\braket{\eta_i(t) \eta_j(t')}$ denoting the two-point correlation function for the classical noise process. Note that $S_k$ are system operators (not necessarily Pauli operators) designating the manner in which the noise affects the system. 

Simulations are performed by Trotterizing $H(t)$ as
\begin{equation}
    U(T)=\mathcal{T}e^{-i\int^{T}_0 H(t) dt} \approx \prod^{N_T}_{k=1} e^{-i H(k\Delta t)\Delta t},
\end{equation}
where $\Delta t$ sets the time resolution of the simulation and $N_T=T/\Delta t$ denotes the number of time steps. In order to examine the effect of correlated noise on the system, one must average over many trajectories of the noise and therefore, perform this full dynamics simulation numerous times.

SchWARMA simulations are performed by partitioning the controlled dynamics and the correlated noise evolution. The total time evolution is composed of products of ideal evolution followed by SchWARMA-generated noise that effectively captures the effect of $H_e(t)$. More concretely, the evolution is explicitly approximated by
\begin{equation}
    U(T)\approx \prod^{N_{sch}}_{j=1} U_E(j\Delta t^\prime)U_{c}(j\Delta t^\prime) U^\dagger_{c}((j-1)\Delta t^\prime),
\label{eq:ad_trotter}
\end{equation}
where $N_{sch}=T/\Delta t^\prime$ is the number of SchWARMA time steps, $\Delta t^\prime = \kappa \Delta t$ is the effective SchWARMA sampling time and
\begin{eqnarray}
    U_{c}(j\Delta t^\prime) &=&U_{c}(j\kappa\Delta t)  \nonumber\\
    &=&  \mathcal{T}e^{-i\int^{j\kappa\Delta t}_0 H_c(t) dt} \approx \prod^{j\kappa}_{k=1} e^{-i H_{c}(k\Delta t)\Delta t}.\quad
\end{eqnarray}
Correlated noise is generated via $U_E(t_j)=\exp(-i \sum_\gamma y^{(\gamma)}_{j} S_\gamma)$, where the variables $y_j^{(\gamma)}$ are defined by the underlying SchWARMA model and thus, they encapsulate the temporal properties of the noise. Simulating the dynamics in this manner allows one to leverage the noiseless approximation of adiabatic dynamics as a base for the faulty simulations. The noisy dynamics are then captured by averaging a desired metric over realizations of Eq.~(\ref{eq:ad_trotter}), where the noise is applied at times $j\Delta t^\prime$ that may be much larger than the Trotter time resolution $\Delta t$.

We consider three different continuously driven systems to demonstrate the efficacy of SchWARMA: (a) a 2-level Landau-Zener system (LZS)~\cite{landau1932:lz,zener1932:lz}, (b) a 3-level LZS~\cite{carroll1986:lz3}, and (c) $N$-qubit adiabatic Grover's Search \cite{roland2002:aqc}. SchWARMA is shown to be in  agreement with full dynamics simulation of each system, while only requiring a fraction of the simulation cost. In the subsequent simulations, $\Delta t/\Delta t'=100$; thus, while the ideal simulation cost remains the same, the cost of a noisy simulation is approximately reduced by a factor of 100.

\begin{figure}[t]
    \centering
    \includegraphics[width=\columnwidth]{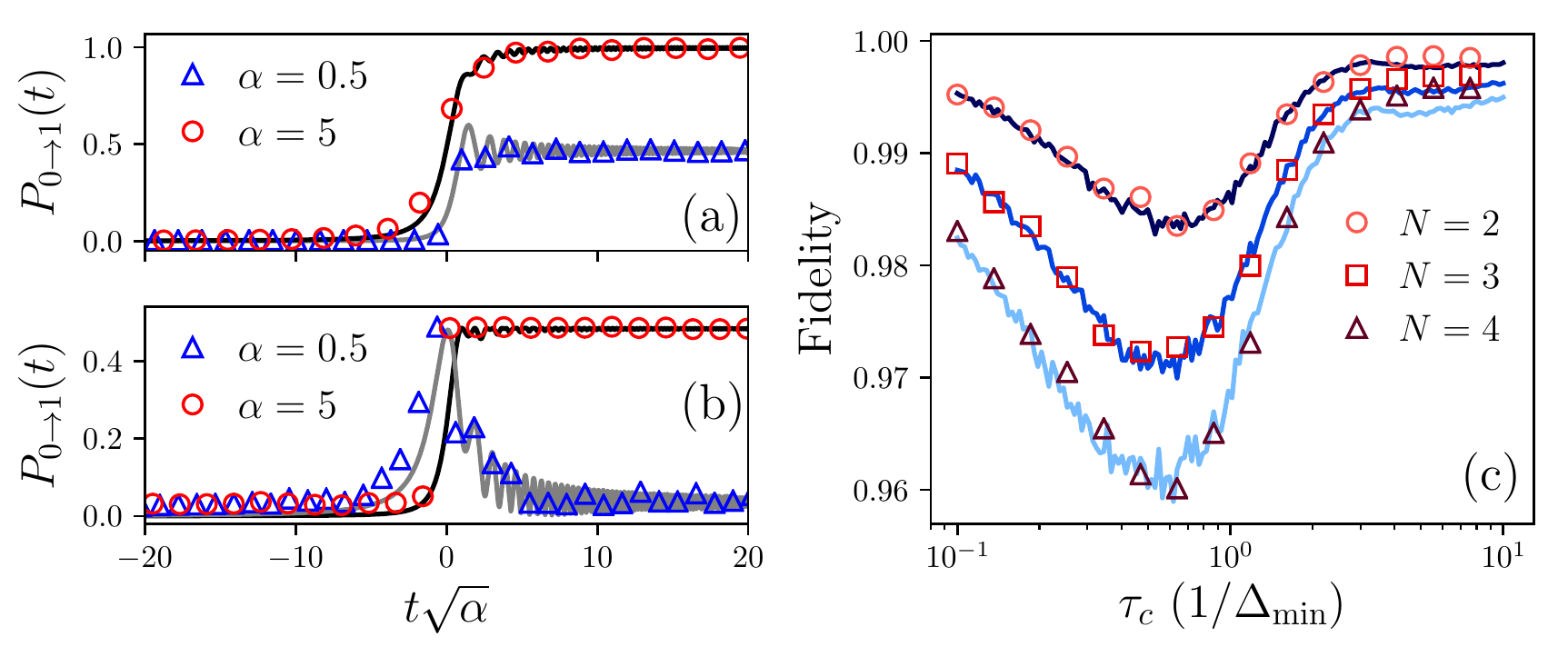}
    \caption{SchWARMA (symbols) vs. analytical results and full dynamics simulation of continuously driven systems (both shown as solid lines). The comparison indicates good agreement between SchWARMA and alternative approaches. LZ results are shown for an (a) 2-level and (b) 3-level system with noise power $f(0)=0.003\alpha$ for different driving rates $\alpha$. Grover's search algorithm comparisons [panel (c)] focus on varying correlation time $\tau_c$ using a fixed noise power of $f(0)=0.001\Delta^2_{\min}$.}
    \label{fig:aqc_schwarma}
\end{figure}

We model the faulty LZ system as 
\begin{equation}
H_{c}(t)=2\Delta S_x + 2\alpha t S_z
\end{equation} 
and $H_e(t)=\eta_x(t)S_x$, where $S_x$ and $S_z$ represent spin operators for either the spin-1/2 or spin-1 system. We simulate the dynamics from $t\ll0$ to $t\gg 0$ to effectively capture the LZ behavior from $t\rightarrow -\infty$ to $t\rightarrow \infty$ and compare our results to the analytical expressions obtained in Ref.~\cite{kenmoe2013:lz} for the transition probability from $\ket{0}$ to $\ket{1}$ in the presence of slow correlated noise. Fig.~\ref{fig:aqc_schwarma} conveys that the analytical expression for transition probability as a function of total time is in good agreement with SchWARMA-driven simulations for both systems (a) and (b). The total time, noise correlation time, and noise power are all normalized with respect to the drive strength $\alpha$ for distinct values of $\lambda=\Delta^2/2\alpha$. The total drive time and noise correlation time are given by $T=T_0/\sqrt{\alpha}$ and $\tau_c=\tau_0/\sqrt{\alpha}$, respectively. The normalized times $T_0=600$ and $\tau_0=100$ are chosen for the simulations shown in Fig.~\ref{fig:aqc_schwarma}. We express the noise power $f(0)=0.003\alpha$ in terms of the driving rate. Parameter definitions are chosen in accordance with Ref.~\cite{kenmoe2013:lz}.

In addition to the LZS, we consider the adiabatic implementation of Grover's search algorithm:
\begin{equation}
    H_{ad}(t)=\alpha(t)[I-\ket{+}\bra{+}] + (1-\alpha(t))[I-\ket{m}\bra{m}],
\end{equation}
where $\alpha(t)$ describes the time-optimal Grover control \cite{roland2002:aqc}, $\ket{+}$ denotes the equal superposition state, and $\ket{m}$ represents the target, marked state. The noise is modeled as $H_E(t)=\eta(t)\sum_i \sigma^z_i$; thus, we consider collective dephasing with respect to the computational basis. Simulations shown in Fig.~\ref{fig:aqc_schwarma} include comparisons between fully Trotterized dynamics and SchWARMA simulations as a function of correlation time with respect to $\Delta_{\min}$, the minimum spectral gap between the ground state and first excited state. The fidelity metric is $F=|\braket{\Phi(T)|\psi(T)}|^2$, where $\ket{\Phi(t)}$ denotes the instantaneous ground state of $H_{ad}(t)$ and $\psi(T)$ is the time-evolved state at total time $T$. The noise power and correlation times are also normalized with respect to $\Delta_{\min}$. The comparisons indicate that SchWARMA is in good agreement with analytical predictions and full dynamics simulations for a variety of $N$-qubit system sizes.

%
\section{Discussion and Future Work}\label{sec:conclusions}
We presented an approach, motivated by classic ARMA modeling, that parameterizes and simulates time-correlated dynamics in quantum systems.  To highlight the versatility of this technique, we demonstrated its applicability in simulating dephasing noise (single and multi-axis) and non-unital dissipative noise, applied to single qubits as well as multi-qubit circuits. The single-qubit (and qutrit) simulations using SchwARMA produce strong agreement with established theory, and the multi-qubit simulations tracked accurately to stochastic Liouville simulations, at substantial computational savings.

We emphasise that SchWARMA is not limited to the examples considered above, or even to channels near the identity as we have done. In particular, semi-classical noisy Hamiltonian dynamics are converted to SchWARMA models by multiplying the Hamiltonian elements by $i$, c.f., the single- and multi-axis dephasing examples above.  The multiplication by $i$ converts the Hermitian Hamiltonians to the skew-symmetric constraint needed for that portion of the Stiefel manifolds tangent space.  This applies to arbitrary $N$-level systems including multi-qubit entangling noise. To extend the approach to arbitrary stochastic noise on Kraus operators, the $\log$ map for the Stiefel manifold \cite{zimmermann2017matrix} can be used.  Here, this corresponds to the standard matrix logarithm on the Kraus-stacked Stinespring form to compute its corresponding Lie algebra element (multiplied by $i$), where the first $N$ columns correspond to the tangent space of the corresponding Stiefel manifold. 

Beyond the application of SchWARMA to general CPTP maps and arbitrary stationary Gaussian noise, the underlying ARMA formalism itself can be further generalized to model additional phenomena. At its core, SchWARMA is itself a non-linear vector ARMA (VARMA) model \cite{fiori2014auto,holan2010arma}, and generalizing the underlying VARMA model can be done in a similar fashion to standard ARMA models, which would then scale the the tangent space elements and be matrix exponentiated.
There is a rich field of generalizations of ARMA models in the literature \cite{holan2010arma}, which may be appropriate for different physical phenomenon than standard ARMA models. For example, ARIMA models nonstationary noise, SARMA and PARMA models are often used to model stochastic variations around some otherwise periodic behaviors, and GARCH models model volatility in the overall noise power \cite{bollerslev1986generalized}. Additional nonlinearities could be used to model non-Gaussian semiclassical noise, such as that found in \cite{norris2016qubit}. We also expect that time-varying model coefficients and cross-correlated driving terms $x_k^{(\ell)}$ will find use in the modeling of control noise and spatial correlations, respectively.

The results presented here open up many avenues for future work. We have shown it to have utility in experimental noise injection \cite{murphy2021universal} and analysis of quantum error correction \cite{clader2021impact}. The ability to generate temporally-correlated errors at a low computational cost can be immediately applied to other simulations and experiments relevant to quantum error correction, such as for threshold analysis in the presence of absolutely continuous errors~\cite{PhysRevA.95.062338}.  Similarly, SchWARMA could be used to analyze and extend techniques for 
quantum system characterization \cite{knill2008randomized,blume2013robust, Merkel2013, greenbaum2015introduction} and quantum noise spectroscopy in the presence of correlated noise. 
Given the connections between 
ARMA  and the fields of spectrum estimation, classical signal processing, and control, we believe there are a number of straightforward generalizations from the classical realm to that of open quantum systems. In particular, this approach could be a ``discrete time'' analogue to the filter function formalism \cite{PhysRevLett.87.270405, PhysRevLett.113.250501}, allowing for noise filtering and mitigation.

\begin{acknowledgments}
We thank Colin Trout, Tim Sweeney, Lorenza Viola, and Leigh Norris for reviewing the manuscript and for fruitful conversations during its preparation. KS, GQ, and PT acknowledge funding from the DOE Office Science contract 4000167418. KS, GQ, PT, and BDC acknowledge funding from the DOE Office of Science Grant Number DE-SC0020316. KS and GQ acknowledge support from ARO MURI grant W911NF-18-1-0218. A portion of this research was supported by the Intelligence Advanced Research Projects Activity via Department of Interior National Business Center contract number 2012- 12050800010. The U.S. Government is authorized to reproduce and distribute reprints for Governmental purposes notwithstanding any copyright annotation thereon. The views and conclusions contained herein are those of the authors and should not be interpreted as necessarily representing the official policies or endorsements, either expressed or implied, of IARPA, DoI/NBC, or the U.S. Government.
\end{acknowledgments}

\bibliography{references.bib}

\appendix


\section{Relationship between SchWARMA and Lindblad Master Equation}\label{app:master}
The Lindblad master equation \cite{gorini1976completely,lindblad1976generators,breuer2002theory} can be viewed as a differential version of a CPTP map under Markovian evolution. In this section we will show the relationship between the Lindblad master equation and dynamics on Stiefel manifolds and thus the SchWARMA approach.
A convenient form of the Lindblad master equation  is
\begin{equation}
    \dot{\rho}=-i[H,\rho]+\sum_{\alpha}\left(L_\alpha \rho L_\alpha^\dagger-
\frac{1}{2}\left\{L_\alpha^\dagger L_\alpha,\rho\right\}\right)
\end{equation}
where $H$ is the system Hamiltonian, and $\{L_\alpha\}$ are the so-called Lindblad operators. We will essentially follow the derivation in \cite{jagadish2019invitation} using the stacked Kraus operator formulation of SchWARMA. In particular, we show that the tangent space elements of the Stiefel manifold are related to the Lindblad generators.

Consider an arbitrary Stiefel manifold tangent space element decomposed into $N\times N$ blocks
\begin{equation}
    X=\begin{bmatrix}A\\ B_1\\\vdots\\ B_{K-1}\end{bmatrix}\,
\end{equation}
recalling that $A$ is skew symmetric like the matrix $-iH$ for an arbitrary Hamiltonian.  The SchWARMA approach utilizes the exponential map on the Stiefel manifold, which corresponds to a $KN\times KN$ unitary evolution of a density matrix consisting of the system and the environment, in the spirit of the Stinespring form of the CPTP map \cite{wood2015tensor}. This is then followed by tracing out the environment to obtain the evolution of the system density matrix. 

Let us start by showing, how an individual SchWARMA step relates to the evolution of density matrices under CPTP maps. The propagation of a joint system-environment state $\rho_{ES}$  is given by  joint unitary dynamics,
\begin{equation}
    \dot{\rho}_{ES} = \left[\left(\begin{matrix} A & -B^\dagger\\B & \mathbf{0}
    \end{matrix}\right),\rho_{ES}\right]\triangleq[iH_{ES},\rho_{ES}]
\end{equation}
where, the initial state of the system+environment is considered to be unentangled, $\rho_{ES}(0)=|0_E\rangle\langle 0_E|\otimes \rho_S(0)$ and $B=[B_1^\dagger,\cdots B_{K-1}^\dagger]^\dagger$. For a SchWARMA gate time $T$, and constant $A$ and $B$, this results in 
$\rho_{ES}(T)=U_{ES}(T)\rho_{ES}(0)U_{ES}(T)^\dagger$ where
\begin{equation}\label{eq:full_unitary}
    U_{ES}(T)=\exp\left(T\begin{bmatrix} A & -B^\dagger\\B & \mathbf{0}
    \end{bmatrix}\right)\,.
\end{equation}
%
The second piece of an individual SchWARMA step is to trace out the environment, so that $\rho_S(T)=\text{Tr}_{E}(\rho_{ES}(T))$ which amounts to Kraus operators
\begin{equation}
    \rho_S(T)=\sum_{k=1}^KM_k(T)\rho_S(0)M_k(T)^\dagger
\end{equation}
where 
\begin{equation}
    M_k(T)=(\langle k_E|\otimes I_N)U_{ES}(T)I_{KN^2,N},
\end{equation}
where $I_N$ is the $N\times N$ identity matrix and $I_{KN^2,N}$ is the first $N$ columns of $I_{KN^2}$.  In other words, the Kraus operators are the $N\times N$ blocks of the first $N$ columns of $U_{ES}$, as desired.

Next, we will analyze these dynamics in the limit of small $T$  to show how they relate to the Lindblad master equation.  For small $\Delta T$, we could say that
\begin{equation}
     U_{ES}(\Delta T)\approx I_{ES}+\Delta T\begin{bmatrix} A &-B_1^\dagger &\dots\\
        B_1&\mathbf{0}\\
        \vdots\end{bmatrix}\,,
\end{equation}
to determine the limiting Kraus operators $M_k(\Delta T)$.
However, this ignores the Stiefel manifold orthonormal column constraint (equivalently, the TP constraint) so we need some correction $B_0$ where 
\begin{equation}
    (I_N+A+B_0)^\dagger(I_N+A+B_0)=I_N-\sum_{k=1}^{K-1} B_k^\dagger B_k\,,
\end{equation}
so 
\begin{equation}
    A^\dagger+A+B_0^\dagger+B_0 + \mathcal{O}(\Delta T^2)=-\sum_{k=1}^{K-1} B_k^\dagger B_k\,.
\end{equation}
note that $A$ is skew-Hermitian (so $A^\dagger+A=0$) and thus 
\begin{equation}
    B_0=-\frac{1}{2}\sum_{k=1}^{K-1}B_k^\dagger B_k\,.
\end{equation} 
This suggests a set of infinitesimal Kraus operators
\begin{equation}
    M_k(\Delta T)\approx \left\{\begin{array}{lr} I_N+A+B_0\,,&k=1\\
    B_{k-1}\,,&2\leq k\leq K\,.\end{array}\right.
\end{equation}

Then, in a similar fashion to \cite{jagadish2019invitation}, our Stiefel manifold construction has (up to order $\Delta T$) 
\begin{equation}
    \rho+\Delta \rho = (I+A+B_0)\rho(I+A+B_0)^\dagger +\sum_{k=1}^{K-1}B_k\rho B_k^\dagger\,.
\end{equation}
Subtracting $\rho$ from both sides and substituting $A=-iH$, we have
\begin{equation}
\Delta \rho = -i[H,\rho]+\{B_0,\rho\}+(A+B_0)\rho (A+B_0)^\dagger +\sum_{k=1}^{K-1}B_k\rho B_k^\dagger\,.
\end{equation}
Note that the term $(A+B_0)\rho(A+B_0)^\dagger$ is second order in $A$ and fourth order in $B_i$, so we have that, up to the original order, 
\begin{align}
\Delta \rho &= -i[H,\rho]+\{B_0,\rho\}+\sum_{k}^KB_i\rho B_i^\dagger\\
&=-i[H,\rho]+\sum_{k=1}^{K}\left(B_k\rho B_k^\dagger-\frac{1}{2}\{B_k^\dagger B_k,\rho\}\right)\,.
\end{align}

This indicates several relevant factors for SchWARMA.  First, to identify tangent space elements that generate target non-unitary Kraus operators (with some parametric form), we need to first find the Kraus operator that can be set closest to $I_N$, then set $B_k$ to the Kraus operators that are not that near-$I_N$ operator.  Equivalently, if we have a known master equation in the above canonical form that we wish to emulate with SchWARMA, the Stiefel manifold tangent elements $B_k$ should be set to $L_\alpha$.  Again, while the Lindblad and SchWARMA approaches are equivalent to first order, SchWARMA temporarily propagates the bath and so the trajectories will diverge.  That said, it is clear that small SchWARMA steps will approximate the Lindblad master equation.  Furthermore, as the SchWARMA approach as formulated always produces CPTP maps (that may arise from non-CP divisible trajectories), a CP-divisible trajectory could be derived from the Lindblad master equation that produces equivalent maps at the gate time-scale.  

This suggests an alternative (and essentially equivalent) usage of SchWARMA for stochastic Lindblad master equations where individual time-correlated steps are defined by $|\rho_{k+1}\rangle\rangle = \mathcal{L}_k|\rho_k\rangle\rangle\,$,
with 
\begin{eqnarray}
\mathcal{L}_k &=& \exp\left[-iI_N\otimes H_k-iH_k\otimes I_N +\sum_\alpha ||y_{k}^{(\alpha)}||^2\right. \times \nonumber\\
&&\times \left.\left(L_\alpha^*\otimes L_\alpha -\frac{1}{2}I\otimes L_\alpha^\dagger L_\alpha -\frac{1}{2}(L_\alpha^\dagger L_\alpha)^\top\otimes I\right)\right],\nonumber\\
\end{eqnarray}
where $H_k$ is a "unitary" SchWARMA Hamiltonian, and $y_k^{(\alpha)}$ are the outputs of (possibly complex) ARMA models as in the original SchWARMA formulation.  This formulation is clearly dissipative and CP-divisible, as the coefficients Lindblad coefficient $||y_{k}^{(\alpha)}||^2$ are nonnegative.  It is clear, however, that maintaining the SchWARMA "bath" offers a potential path forward for non-CP divisible dynamics.

\section{Determining SchWARMA Model Coefficients}\label{app:coeffs}
The power spectrum of the noise generated by a SchWARMA model is determined by the underlying ARMA model coefficients.  Furthermore, ARMA models are dense in the space of valid power spectra, meaning that ARMA (as well as AR or MA) coefficients exist that allow for SchWARMA approximations of \text{any} power spectra, to any desired accuracy.  Digital signal processing has a number of methods for determining ARMA coefficients, referred to in this domain as \textit{filter design}. Band-limited (including low-pass, band-pass, high-pass, and multi-band) spectra can be parametrically  generated using window-based methods, with the choice of window determining  features such as flatness in the pass-bands, roll-off rate to stop-bands, and overall suppression in the stop-bands \cite{oppenheim1999discrete}.  Other options for filter design can produce (approximate) fits for arbitrary given spectra, such as the Parks-McClellan algorithm\cite{parks1972chebyshev} or convex optimization \cite{wu1996fir}.  Specific techniques exist for certain spectra, such as $f^{-\alpha}$ noise \cite{plaszczynski2007generating} or line spectra \cite[Ch.~4]{stoica2005spectral}.

In this work, we are also faced with the challenge of taking a parametrically defined noise spectrum at a fine time scale (the Trotter ``step'' of our stochastic master equation) and computing an equivalent effective noise at a coarser time scale (the ``gate'' time scale of SchWARMA).  We assume that both noises are Gaussian and wide-sense stationary with discrete-time autocovariance functions $r_f[k]$ and $r_s[k]$ for the ``fast'' master equation time scale and the ``slow'' SchWARMA time scale, respectively.  Supposing we wish to design a SchWARMA model that effectively captures the noise at an integer factor $T$ slower, it is not sufficient to set $r_s[k]=r_f[kT]$, as this ignores the intuition that the fast noise should be in some sense integrated (i.e., added) during the $T$ steps of the faster model, as in a random walk.  Thus, we need to relate $r_s[k]$ to the sum of $T$ steps of the fast model. To achieve this, we exploit the well-known relationship about the variance of the sum of $N_g$ correlated Gaussians $X_i$:
\begin{equation}
    \text{Var}\left(\sum_{i=1}^{N_g} X_i\right)=\sum_{i=1}^{N_g}\text{Var}(X_i)+2\sum_{1\leq i<j\leq N_g}\text{Cov}(X_i,X_j)\,.
\end{equation}
Given a target fast time scale noise autocovariance $r_f$ we can set up a system of $P$ equations to solve for the unknown $r_s$ using
%
\begin{widetext}
\begin{align}
\label{eq:conv1}r_s[0] &= Tr_f[0]+2\sum_{i=1}^{T-1}(T-i)r_f[i]\,,\\
2r_s[0]+2r_s[1] &= 2Tr_f[0]+2\sum_{i=1}^{2T-1}(2T-i)r_f[i]\,,\\ \nonumber
& \vdots \quad \textnormal{$N_g$ times} \\ 
\label{eq:conv2}N_g r_s[0]+2\sum_{j=1}^{N_g-1}(N_g-j)r_s[j]&=N_gTr_f[0]+2\sum_{i=1}^{N_gT-1}(N_gT-i)r_f[i]\,,
\end{align}
\end{widetext}
%
where we have substituted e.g., $r_f[|i-j|]$ for $\text{Cov}(X_i,X_j)$ per the definition of wide sense stationary.  Note that the left hand side of the equations correspond to the variance of the sum of the slow time process, and the right hand side is the variance of the sum of the fast time process sampled every $T$ steps.  Next, we then use the spectral factorization approach to filter design outlined in \cite{wu1996fir} to generate coefficients for a SchWARMA model.

\section{Error Analysis}\label{app:error}
%

In this section, we analyze the different sources of error introduced to the simulation from the discrete time-step of the SchWARMA model that is at a coarser time scale compared to the correlation time of native noise. First, we consider the case where the noise (and control) commute at all times. In this case, we show that the SchWARMA model can be chosen to have the desired accuracy in comparison to the exact simulation at this discrete time-scale. Next we consider the more general scenario of non-commuting noise and applied controls. In this case, utilizing the Magnus expansion, we show that the SchWARMA model introduce an error proportional to the power of the modeled noise. Finally, we consider what is actually the dominant error source in a practical application of  SchWARMA (or other Trotterized stochastic master equation approach), namely the sampling error introduced by Monte Carlo simulation. This is precisely what is observed in the absolute error seen in  circuit simulations in the main text. 

\subsection{Commuting Hamiltonians}

Consider the case of the noisy Hamiltonian $H(t)=\eta(t)\sigma$, for some bounded Hermitian operator $\sigma$. Let $\eta(t)=\mu(t)+\tilde{\eta}(t)$ with $\mu(t)$ a deterministic integrable function and $\tilde{\eta}(t)$  a stochastic process satisfying $E[\tilde{\eta}(t)]=0$ and $E[\tilde{\eta}(t)\tilde{\eta}(s)]\leq C$, i.e., uniformly bounded for all $s,t\in\mathbb{R}$. Examples of such processes include single axis dephasing evolutions in the absence of control, or single axis multiplicative control noise. Then, for any time $T>0$ and a specific stochastic trajectory of $\eta$, $U(T)=\exp(-i\int_0^Tdt\,\eta(t)\sigma)$.  Thus, the expected superoperator $\bar{\mathcal{L}}(T)=E\left[U(T)^*\otimes U(T)\right]$.  Since $\sigma$ is Hermitian, it can be diagonalized $\sigma=\Sigma\Lambda\Sigma^\dagger$, so that $U(T)=\exp(-i\int(\eta(t)\sigma))=\Sigma\exp(-i\int\eta(t)\Lambda)\Sigma^\dagger$, and furthermore, the Liouvillian can be diagonalized by
\begin{widetext}
\begin{equation}
    \mathcal{L}(T)=\Biggl(\Sigma^*\otimes\Sigma\Biggr)\Biggl(\exp\left(i\int_0^Tdt\, \eta(t)\Lambda^*\right)\otimes\exp\left(-i\int_0^Tdt\,\eta(t)\Lambda\right)\Biggr)\Biggl(\Sigma^\top\otimes\Sigma^\dagger\Biggr)
\end{equation}
\end{widetext}
and so the dynamics of $\mathcal{L}$ are ultimately determined by $\exp(i\int\eta(t)(\lambda_j-\lambda_k))$ where $\lambda_j$, $\lambda_k$ are eigenvalues of $\sigma$.  Thus, by averaging over random trajectories, we have that $\bar{\mathcal{L}}(T)$ is determined by the expectations $E[\exp(i\int_0^Tdt\,\eta(t)(\lambda_j-\lambda_k)]$.  Now since $\eta(t)$ is a stochastic process, $\int_0^Tdt\,\eta(t)\triangleq S(T)$ is a random variable and the $E[\exp(iS(T)(\lambda_j-\lambda_k))]$ is its characteristic function evaluated at $(\lambda_j-\lambda_k)$, which exists.
Now, if we assume that $\eta(t)$ is a member of a closed (under addition) functional family, such as a Levy $\alpha$-stable distribution (an example of such a distribution is the Gaussian distribution), then $S(T)$ are all the same form of random variable (e.g., Cauchy, Gaussian), and we can exploit known results of characteristic functions.

In the case that $\eta$ (and thus $S$) are Gaussian, we have that 
\begin{align}
\label{seq:expiS}
    & E[\exp(i S(T))(\lambda_j-\lambda_k))] \\ \nonumber
    &=E[\exp(iS(T)(\lambda_j-\lambda_k))]\\ \nonumber
    &=\exp\left(i(\lambda_j-\lambda_k)E[S(T)]\right) \\ \nonumber
    & \times \exp\left(\frac{1}{2}\text{Var}(S(T))((\lambda_j-\lambda_k)^2)\right)
\end{align}
By construction, we have that $E[S(T)]=\int_0^T\,dt \mu(t)$ which exists by assumption, and 
$\text{Var}(S(T))=E[S(T)^2]-E[S(T)]^2$ with
\begin{widetext}
\begin{equation}
    \begin{aligned}
    E[S(T)^2] &= E\left[\left(\int dt\, \mu(t)+\tilde{\eta}(t)\right)^2\right]\\
    &= E\left[\left(\int_0^T dt\,\mu(t)+\tilde{\eta}(t)\right)\left(\int_0^T ds\, \mu(s)+\tilde{\eta}(s)\right)\right]\\
    &= E\left[\int_0^T dt\, \mu(t)\int_0^Tds\,\mu(s)\right] +2E\left[\int_0^T dt\,\mu(t)\int_0^T ds\,\tilde{\eta}(s)\right] +E\left[\int_0^Tdt\,\int_0^Tds\,\tilde{\eta}(t)\tilde{\eta}(s)\right]\\
    &=E[S(T)]^2+E[S(T)]\int_0^T ds\,E[\tilde{\eta}(s)]+\int_0^T dt\,\int_0^T ds\, E[\tilde{\eta}(t)\tilde{\eta}(s)]
    \end{aligned}
\end{equation}
\end{widetext}
since, by assumption $E[\tilde{\eta}(s)]=0$, we have that
$\text{Var}(S(T))=\int_0^T dt\,\int_0^T ds\, E[\tilde{\eta}(t)\tilde{\eta}(s)]$ which depends entirely on the two point correlation function of $\tilde{\eta}$.  Under the additional assumption that $\tilde{\eta}$ is wide sense stationary, then $E[\tilde{\eta}(t)\tilde{\eta}(s)]=E[\tilde{\eta}(|s-t|)\tilde{\eta}(0)]$, and we can define the autocovariance function $r_{\tilde{\eta}}(t)=E[\tilde{\eta}(|t|)\tilde{\eta}(0)]$ and thus
\begin{equation}\label{eq:ctsvar}
    \begin{aligned}
    \text{Var}(S(T))&=\int_{0}^{T}dt\,\int_{0}^T\,ds\,r_{\tilde{\eta}}(t-s)\\
    &=\int_{-T}^T dt\,r_\eta(t)(T-|t|)\\
    &=\frac{T^2}{\sqrt{2\pi}}\int_{-\infty}^\infty d\omega\, R_{\tilde{\eta}}(\omega)\sinc^2\left(\frac{T\omega}{2\pi}\right)
    \end{aligned}
\end{equation}
where $R_{\tilde{\eta}}(\omega)=\frac{1}{\sqrt{2\pi}}\int_{-\infty}^\infty dt\, r_{\tilde{\eta}}(t)e^{-i\omega t}$ is the continuous time Fourier transform of the autocorrelation of $\tilde{\eta}$, i.e., the power spectrum of $\tilde{\eta}$.

The above expression is useful for determining SchWARMA models for dephasing noise, as it indicates that if we are interested in discrete times $kT$ (e.g., as in a circuit with fixed gate time), then we need to find an ARMA model with autocovariance $r_s[k]$ that satisfies the continuous time version of Eq.~\ref{eq:conv1}-\ref{eq:conv2}, i.e., replacing the summations on the right hand side of Eq.~\ref{eq:conv1}-\ref{eq:conv2} with integrals as in Eq.~\ref{eq:ctsvar}.  Again, we reiterate that since the possible power spectra of  AR, MA, and ARMA models are each dense in the space of valid power spectra (and thus valid autocovariances), finding an AR, MA, or ARMA model that is arbitrarily close to given autocovariance is \textit{always} possible in principle, and many practical methods for finding solutions exist such as \cite{wu1996fir}. Again, we emphasize that this is a matching on the discrete $kT$ timescale of interest, and that the continuous time power spectrum of a derived ARMA model will always be periodic in $1/(2T)$, but result in equivalent sampled dynamics at time steps $kT$, meaning this difference is irrelevant on the gate time scale.

The above discussion highlights a subtle distinction about conversion between continuous and discrete time-scales, or between different discrete-time scales.  Clearly, a given functional form for an autocovariance or power spectrum, such as squared exponential, is not preserved by the conversion due to the time-integration. This agrees with the intuition that a noise source with correlation much shorter than a gate time should be well modeled by white noise at the gate time scale.  That is not to say, however, that the short-time correlations are completely ignored by a SchWARMA model, they are instead aggregated at the coarser time scale (and to the accuracy of the ARMA fit). This also means that parameter sweeps at the gate time scale do not map perfectly to parameter sweeps at the continuous time scale, but coarser features such as correlation time, will remain accurate to the order of the discretization.

\subsection{Average error operator after a single gate }


In this section, we consider a single gate, $U_c$ that is applied over a time $T$ in the presence of a native noise process.  Using the Magnus expansion, we derive the expression for an average error operator applied \emph{after} the perfect gate to model the dynamics. This average error operator can be then reduced to a SchWARMA model to characterize the error induced by the noise at this gate time-scale.

Consider the gate $U_c(0,T)=\mathcal{T}\left[\exp\left(\int_0^T dt H_c(t)\right)\right]$ being applied by the Hamiltonian $H_c$ for time $T$, with an unwanted noise process given by the Hamiltonian $H_\eta (t)$. We can define the control and noise Hamiltonians by a given set of basis operators  $\{\boldsymbol{P}_p\}$,
\begin{align}
    H_\eta=\sum_p \eta_p(t) \boldsymbol{P}_p,\quad H_c(t)=\sum_p h_p(t) \boldsymbol{P}_p.
\end{align}
For example, for multi-qubit states $\boldsymbol{P}_p$ could be given by a tensor product of Pauli matrices. The $\eta_p(t)$ describes the native noise process along the given basis state $\boldsymbol{P}_p$. We have the exact time-evolution operator as a time-ordered product,
\begin{align}
    U(0,T)&=\mathcal{T}\left[\exp\left(\int_0^T dt\  H_c(t)+H_\eta(t)\right)\right]\\
    &=\tilde{U}(0,T) U_c(0,T)= U_c(0,T)\tilde{U}^\prime(0,T)
\end{align}
where, $\tilde{U}^\prime(t)= U_c^\dagger(0,t)\tilde{U}(0,t) U_c(0,t)$. The operator, $\tilde{U}(0,T)$ characterizes the error in the application of this gate. In the following, we use the Magnus expansion to obtain an approximate expression for the error applied.

Going to an interaction frame, using $idU/dt=(H_c+H_\eta)U$, and $-idU_c^\dagger/dt=U_c^\dagger H_c$, we have the following expression for $\tilde{U}^\prime$, 
\begin{align}
    i\frac{d\tilde{U}^\prime}{dt}&=iU_c^\dagger\frac{dU}{dt}+i\frac{dU_c^\dagger}{dt}U=(U_c^\dagger H_\eta U_c)\tilde{U}^\prime=\tilde{H}_\eta \tilde{U}^\prime
\end{align}
for which we have the standard Magnus expansion,
\begin{align}
    \tilde{U}^\prime(0,T)&=\exp\left[-i\sum_{n=1}^\infty\Phi_n(T)\right],
\end{align}
with
\begin{align}
    \Phi_1(T) &=\int_0^T dt \ U_c^\dagger(0,t) H_\eta(t)U_c(0,t) \\ \nonumber
    & =\int_0^T dt \ \tilde{H}_\eta(t),\\
     \Phi_2(T) &=-\frac{i}{2}\int_0^T dt_1\int_0^{t_1} dt_2 \ \left[\tilde{H}_\eta(t_1),\tilde{H}_\eta(t_2)\right],
\end{align}
and so on. The expression for $\tilde{U}(T)$ is
\begin{align}
    \tilde{U}(0,T)&=U_c(0,T)\exp\left[-i\sum_n\Phi_n(T)\right]U_c^\dagger(0,T)\\
    &=\exp\left[-i\sum_n U_c(0,T)\Phi_n(T)U_c^\dagger(0,T)\right]
\end{align}
 We can simplify the above expression using the basis elements $\{\boldsymbol{P}_p\}$, Defining,
\begin{align}
U_c(t,T) \boldsymbol{P}_pU_c^\dagger(t,T)=\sum_{p}c_{pq}(t)\boldsymbol{P}_q,
\end{align}
we obtain the following expressions for first two terms in the Magnus expansion,
\begin{eqnarray}
  U_c(0,T)\Phi_1(T)U_c^\dagger(0,T) &=& \int_0^T dt\ U_c(t,T)H_\eta(t)U_c^\dagger(t,T) \nonumber\\
  &=& \sum_{pq}\left(\int_0^T dt\ \eta_p(t) c_{pq}(t)\right)\boldsymbol{P}_q,\nonumber\\
\end{eqnarray}
\begin{widetext}
\begin{align}
     U_c(0,T)\Phi_2(T)U_c^\dagger(0,T)&=-\frac{i}{2}\sum_{p_1q_1 p_2 q_2}\int_0^T dt_1\int_0^{t_1} dt_2 \ \eta_{p_1}(t_1)\eta_{p_2}(t_2)c_{p_1q_1}(t_1)c_{p_2q_2}(t_2)\left[\boldsymbol{P}_{q_1},\boldsymbol{P}_{q_2}\right]
\end{align}
\end{widetext}

In the following analysis, we truncate the Magnus expansion, which is a reasonable assumption to make in the limit of weak noise, $T\max_t\{\|H_\eta(t)\|\}\ll 1$. Truncating at the first term in the Magnus expansion, we have the following expression for the error operator,
\begin{align}
    \tilde{U}(0,T) & \approx\tilde{U}_1(0,T) \\ \nonumber
    & = \exp\left[-iU_c(0,T)\Phi_1(T)U_c^\dagger(0,T)\right], \\ \nonumber
    & = \exp\left[-i\sum_{pq}\left(\int_0^T dt\ \eta_p(t) c_{pq}(t)\right)\boldsymbol{P}_q\right] \nonumber \label{seq:magnus-error_first}\\
    & = \exp\left[-i\sum_{pq}\left(\int_0^T dt\ s_q(t)\right)\boldsymbol{P}_q\right] \\ \nonumber 
    & =\exp\left[-i\sum_{q} S_q(T)\boldsymbol{P}_q\right]\,,\label{seq:magnus-error_firstb}
\end{align}
where, we introduce the notation, $S_q(T)=\int_0^T dt\ s_q(t)=\int_0^T dt\ \eta_p(t) c_{pq}(t)$. Note that in principle, the above expression can be generalized to take into account higher order terms in the Magnus expansion, but we leave that for future work. Utilizing \eqref{seq:magnus-error_firstb} in the above expression, one can obtain expressions analogous to \eqref{seq:expiS} to derive the effective noise model for this quantum  operation.

The error in performing the truncation to first  order in the Magnus expansion can be estimated from the norm of the second term in the expansion (this is appropriate again in the limit of weak noise),
\begin{align}
        & \|\tilde{U}(T)-\tilde{U}_1(T)\|\sim \|\Phi_2(T)\| \\ \nonumber
        & \leq \frac{T^2}{4}\sum_{q_1q_2} \left(\max_{t_1,t_2}\{s_{q_1}(t_1)s_{q_2}(t_2)\}\right)\left\|\left[\boldsymbol{P}_{q_1},\boldsymbol{P}_{q_2}\right]\right\|
        \label{seq:magnus-error_expanded}
\end{align}
 Furthermore, the expression in \eqref{seq:magnus-error_expanded} suggests that this approximation is accurate to the cross-covariance of the noise sources. For noise sources that act independently, $E\left[s_{q_1}(t_1) s_{q_2}(t_2)\right]\sim \delta_{q_1q_2}$ one needs to consider higher order  contributions and thus the error scales atleast as the covariance squared. Since the infidelity of the map is proportional to the variance in this case, the SchWARMA model error here will be proportional to the infidelity squared.  This is consistent with other results derived from filter functions, e.g., \cite{ball2015walsh}.

%

\subsection{Generalization to discrete steps}
Let us consider the case where we now have an evolution from $t=0$ to $t=N_gT$, and we are interested in the discrete time-steps $kT$. In this case, one can generalize the derivation, by examining the evolution from time $kT$ to $(k+1)T$,
\begin{align}
    U(kT,(k+1)T)&=\mathcal{T}\left[\exp\left(\int_{kT}^{(k+1)T} dt\  H_c(t)+H_\eta(t)\right)\right] \\ \nonumber
    & =\tilde{U}(kT,(k+1)T) U_c(kT,(k+1)T) \\ \nonumber
    & = U_c(kT,(k+1)T)\tilde{U}^\prime(kT,(k+1)T)
\end{align}
With this definition, we have the following total evolution operator,
\begin{align}
    U(0,N_gT) & =\tilde{U}((N_g-1)T,N_gT)U_c((N_g-1)T,N_gT) \\ \nonumber
    & \cdots\tilde{U}(T,2T)U_c(T,2T)\tilde{U}(0,T)U_c(0,T)
\end{align}
Here, we have the definition for $\tilde{U}$,
\begin{align}
    \tilde{U}(kT,(k+1)T) &= U_c(kT,(k+1)T) \\ \nonumber
    & \times \exp\left[-i\sum_{n=1}^\infty\Phi_n(kT,(k+1)T)\right] \\ \nonumber
    & \times U_c^\dagger(kT,(k+1)T)
\end{align}
with,
\begin{align}
    \Phi_1(kT,(k+1)T) &=\int_{kT}^{(k+1)T} dt \ U_c^\dagger(kT,t) H_\eta(t)U_c(kT,t) \\ \nonumber
    & =\int_{kT}^{(k+1)T} dt \ \tilde{H}^{(k)}_\eta(t),
\end{align}
etc. Now we can define the transformation,
\begin{align}
U_c(t,(k+1)T) \boldsymbol{P}_iU_c^\dagger(t,(k+1)T)=\sum_{j}c^{(k)}_{ij}(t)\boldsymbol{P}_j,
\end{align}
Typically, $U_c$ is chosen from a list of standard gates, so $c^{(k)}_{ij}$ can be calculated in advance depending on the knowledge of the gates in the circuit. 

Now, we can rewrite by truncating at first order,
\begin{align}
    & \tilde{U}(kT,(k+1)T) \\ \nonumber
    & \approx \exp\left[-i\sum_{pq}\left(\int_{kT}^{(k+1)T} dt\ \eta_p(t) c_{pq}(t)\right)\boldsymbol{P}_q\right]\label{seq:magnus-error}\\ \nonumber 
    &=\exp\left[-i\sum_{q} S^{(k)}_q(T)\boldsymbol{P}_q\right]
\end{align}
Therefore, $S^{(k)}_q(T)$ defines the error model for the discrete time-steps $kT$. We can obtain the properties of the effective error model. We have the following expression for the covariance of the effective noise model,
\begin{widetext}
\begin{align}
    E\left[ S^{(k)}_q(T) S^{(k^\prime)}_{q^\prime}(T)\right]=\sum_{p,p^\prime}\int_{kT}^{(k+1)T} dt\int_{k^\prime T}^{(k^\prime+1)T} dt^\prime\ E\left[\eta_p(t)\eta_{p^\prime}(t^\prime)\right] c^{(k)}_{pq}(t) \  c^{(k^\prime)}_{p^\prime q^\prime}(t^\prime)
\end{align} 
\end{widetext}
So, for example, for the case of uncorrelated and identical stationary noise sources, $E\left[\eta_p(t)\eta_{p^\prime}(t^\prime)\right]=\delta_{pp^\prime}r(t-t^\prime)$,
\begin{widetext}
\begin{align}
     E\left[ S^{(k)}_q(T) S^{(k^\prime)}_{q^\prime}(T)\right]=\int_{kT}^{(k+1)T} dt\int_{k^\prime T}^{(k^\prime+1)T} dt^\prime\ r(|t-t^\prime|) \left[C_k^\top(t^\prime)C_{k^\prime}(t)\right]_{qq^\prime}
\end{align} 
\end{widetext}
where $[C_k(t)]_{ij}=c^{(k)}_{ij}(t)$. This expression closely resembles the standard filter function formalism, albeit expressed in the time instead of frequency domain.  In particular, it illustrates how the sub-gate time-scale dynamics interact with the noise.  In principle, we could further adapt expressions like \eqref{eq:ctsvar} to more accurately represent the ``filtered response'' of the pulse-level control on the noise, and adapt gate-dependent SchWARMA models to accommodate this.  However, in the simulations provided in the main text, we do not account for this effect, i.e., the $C_k$ is approximated as an identity matrix.
Again, as above, \eqref{seq:magnus-error} suggests the SchWARMA approximation error is proportional to the cross-covariance of the noise sources or the noise power squared for independent noise sources.

\subsection{Monte Carlo sampling error analysis}\label{app:error:mc}
\begin{figure}[ht!]
    \centering
    \includegraphics[width=\columnwidth]{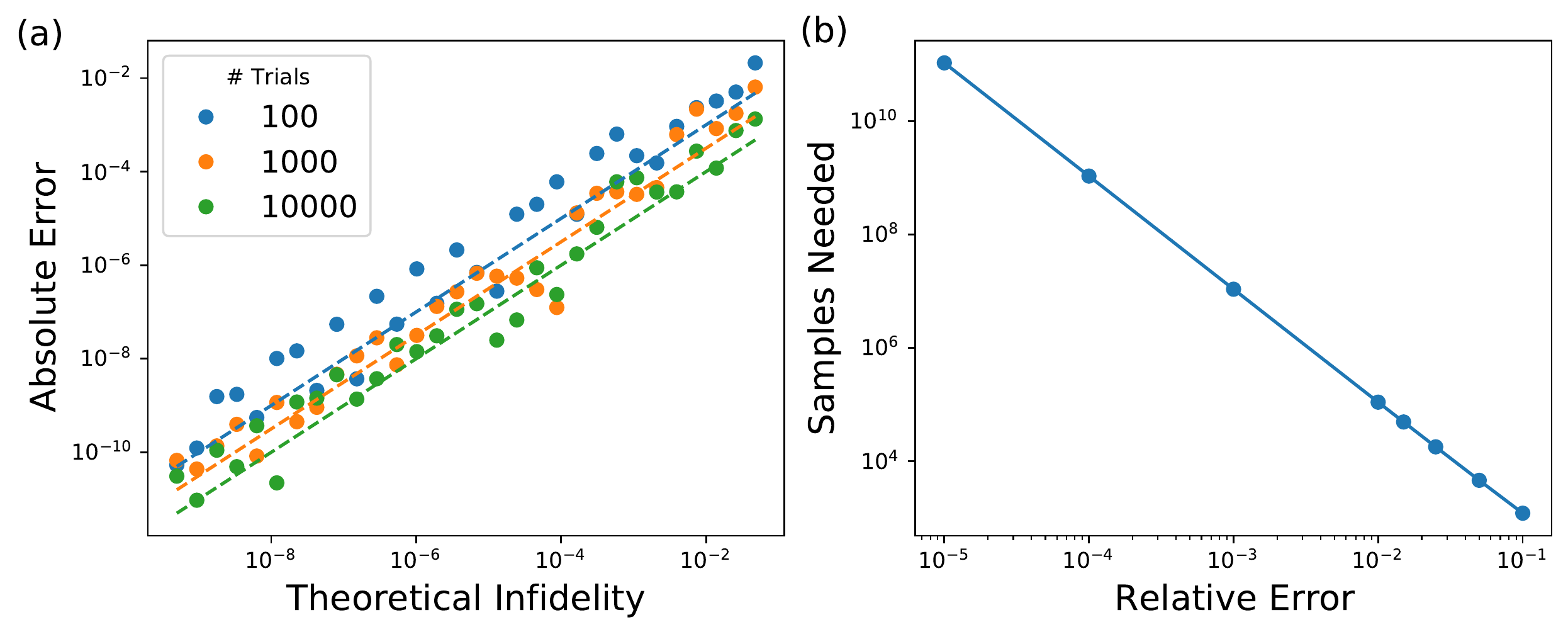}
    \caption{Monte Carlo sampling error analysis. (a) Plots of the absolute error between two SchWARMA Monte Carlo samples with identical theoretical infidelities for a range of infidelities and sample sizes. Note the considerable variation and overlap between the data corresponding to different sample sizes, and the similarity to Fig.~3, right panel. The dashed lines indicate the expected error based on the sample size.  (b) Plot showing the
    required sample size needed for a relative error to achieve a statistically significant result using a single-sided $F$-test at the $p=0.05$ level.  Note the slope of this line is $-2$, in line with the intuition that one needs samples proportional to the square of the sensitivity of the test. 
    }
    \label{fig:mc_analysis}
\end{figure}

Figure \ref{fig:surface_code_noise} in the main text shows that the difference between the SchWARMA and full Trotter simulations is proportional to the infidelity of the noisy circuit. We found that the errors were well within the notional Monte Carlo sampling error, given 1000 Monte Carlo samples data point.
To further illustrate that the error between the SchWARMA and full Trotter simulation results are primarily due to Monte Carlo sampling error, we sampled from identically defined Gaussian random variables of different variances and converted these to $Z$-dephasing errors.  The average process fidelities of these samples was computed and the absolute error between them are shown in Fig.~\ref{fig:mc_analysis}. Following the discussion earlier in this section, the theoretical average process fidelity in this case is directly determined by the variance of the Gaussian distribution.  Note that while the underlying distributions are identical, the empirical averages will be different, and this absolute error will be in proportional to both the the theoretical variance and $||N_{MC}||^{-2}$, where $N_{MC}$ is the sample size.

To assess the statistical significance of the observed differences between process fidelities, we can turn to the theory of $F$-tests \cite{snedecor1980statistical}.  In the analysis of variance, the $F$-test assumes that both samples are Gaussian, and uses the ratio of the sample variances as a test statistic to reject the null hypothesis that the two distributions have the same variance.  While the $F$-test is known to be very sensitive to non-normality in the data, this test remains appropriate for wrapped Gaussians in the weak noise regimes (process infidelity $<0.15$) considered here \cite{mardia2009directional}. By inverting the $F$ distribution as function of $N_{MC}$ we can determine the the number of samples needed to achieve a statistically significant result for a given relative error, see Fig.~\ref{fig:mc_analysis}.  This indicates that our relative errors of $<2.5$\% are well within the numerical sampling error, which would require $N_{MC}>17,500$ for our largest relative errors and $\approx 110,000$ samples for our smallest error. This is already computationally prohibitive for any sort of exploratory analysis, and scales beyond any sort of practical computational budget for threshold computations with independent SchWARMA models, where the analysis above indicates the relative error would be on the order of the infidelity squared (i.e. $<10^{-6}$, for common thresholds).

\end{document}